\documentclass[11pt]{article}

\usepackage{bm}

\usepackage{amsmath,amsthm,amssymb}
\usepackage{enumerate}
\usepackage{graphicx}
\usepackage{epsfig}

\usepackage{hyperref}
\hypersetup{
    colorlinks=true,
    linkcolor=blue,
    filecolor=magenta,      
    urlcolor=blue,
}

\usepackage{enumitem}
\usepackage{hyperref}

  \usepackage{geometry}
\newcommand*{\mailto}[1]{\href{mailto:#1}{\nolinkurl{#1}}}

\newtheorem{theorem}{Теорема}[section]
\newtheorem{definition}[theorem]{Definition}
\newtheorem{lemma}[theorem]{Lemma}
\newtheorem{example}[theorem]{Example}
\newtheorem{proposition}[theorem]{Proposition}
\newtheorem{corollary}[theorem]{Corollary}

\newtheorem{remark}[theorem]{Remark}
\newtheorem{remarks}[theorem]{Remarks}

\DeclareMathOperator{\supp}{supp}
\newcommand{\fr}{\frac}

\newcommand{\Lra}{\Leftrightarrow}

\newcommand{\R}{{\mathbb R}}

\newcommand{\Co}{{\mathbb C}}

\newcommand{\Lam}{{\Lambda}}

\newcommand{\cT}{{\mathcal T}}

\newcommand{\bA}{{\bf A}}
\newcommand{\bB}{{\bf B}}
\newcommand{\cE}{{\cal E}}
\newcommand{\bE}{{\bf E}}

\newcommand{\cH}{{\cal H}}

\newcommand{\bj}{{\bf j}}

\newcommand{\cO}{{\cal O}}

\newcommand{\cm}{{\rm m}}

\newcommand{\al}{\alpha}

\newcommand{\om}{\omega}
\newcommand{\vp}{\varphi}

\newcommand{\si}{\sigma}
\newcommand{\ext}{{\rm ext}}
\newcommand{\De}{\Delta}
\newcommand{\de}{\delta}

\newcommand{\ga}{\gamma}

\newcommand{\ve}{\varepsilon}

\newcommand{\lam}{\lambda}
\newcommand{\Om}{\Omega}

\newcommand{\ti}{\tilde}
\newcommand{\na}{\nabla}
\newcommand{\pa}{\partial}
\newcommand{\rot}{{\rm rot\5}}
\newcommand{\dv}{{\rm div\5}}
\newcommand{\const}{{\rm const}}

\newcommand{\rRe}{{\rm Re\5}}

\newcommand{\ov}{\overline}

\newcommand{\5}{{\hspace{0.5mm}}}

\newcommand{\ds}{\displaystyle}

\date{}

\numberwithin{equation}{section}

\newcommand{\ci}{\cite}
\newcommand{\la}{\label}

\newcommand{\be}{\begin{equation}}
 \newcommand{\ee}{\end{equation}}

 \newcommand{\beqn}{\begin{eqnarray}}
 \newcommand{\eeqn}{\end{eqnarray}}

\newcommand{\ba}{\begin{array}}
 \newcommand{\ea}{\end{array}}

\newcommand{\bd}{\begin{definition}}
 \newcommand{\ed}{\end{definition}}
\newcommand{\bt}{\begin{theorem}}
 \newcommand{\et}{\end{theorem}}

\newcommand{\bp}{\begin{proposition}}
 \newcommand{\ep}{\end{proposition}}

\newcommand{\bl}{\begin{lemma}}
 \newcommand{\el}{\end{lemma}}
\newcommand{\bc}{\begin{corollary}}
 \newcommand{\ec}{\end{corollary}}

\newcommand{\bex}{\begin{example}}
 \newcommand{\eex}{\end{example}}
 
\newcommand{\bexs}{\begin{examples}}
 \newcommand{\eexs}{\end{examples}}

\newcommand{\bexe}{\begin{exercice}}
 \newcommand{\eexe}{\end{exercice}}

\newcommand{\br}{\begin{remark} }
 \newcommand{\er}{\end{remark}}
\newcommand{\brs}{\begin{remarks}}
 \newcommand{\ers}{\end{remarks}}

\newcommand{\bce}{\begin{center}}
\newcommand{\ece}{\end{center}}

\date{}

\numberwithin{equation}{section}

\begin{document}

\bce
{\huge\bf  Lectures on Quantum Mechanics 
\bigskip\\
for mathematicians 
}
 \bigskip \bigskip

 {\Large A.I. Komech} \footnote{
 The research supported by the Austrian Science Fund  (FWF)
 under Grant   No. P28152-N35}
 \medskip
 \\
{\it
  \centerline{Faculty of Mathematics of Vienna University}
 \centerline{Institute for Information Transmission Problems of RAS, Moscow}
      \centerline{Department Mechanics and Mathematics of Moscow State University (Lomonosov)}
 }
  \centerline{alexander.komech@univie.ac.at}

 \bigskip\bigskip
\ece
\bigskip


\begin{abstract}
   The main goal of these lectures -- introduction to 
 Quantum Mechanics
for mathematically-minded readers. The second goal is to discuss the 
mathematical interpretation of the main quantum postulates:
transitions between quantum stationary orbits,
wave-particle duality and probabilistic interpretation.

We suggest a dynamical interpretation
of these phenomena based on the new conjectures on attractors
of nonlinear Hamiltonian partial differential equations.
This conjecture is confirmed 
for a list of {\it model Hamiltonian nonlinear} PDEs
by the results obtained since 1990 
(we survey  sketchy  these results).
However, for the Maxwell--Schr\"odinger equations this conjecture is still an {\it open problem}.

We calculate 
the diffraction amplitude for the scattering of electron beams 
and Aharonov--Bohm shift via  the Kirchhoff approximation.

  
  \end{abstract}
   \bigskip
     


\bigskip\bigskip

\tableofcontents


\section{Introduction}

The first goal of these lectures is to give an introduction to 
Quantum Mechanics
for mathematically-minded readers. The second goal is to discuss the 
mathematical interpretation of main quantum postulates:
 \smallskip\\
I. Transitions between quantum stationary orbits (Bohr 1913).
  \smallskip\\
II. Wave-particle duality (de Broglie 1923).
  \smallskip\\
III. Probabilistic interpretation (Born 1927).
  \smallskip\\
These postulates were inspired by  empirical observations. However, after the discovery of the Schr\"odinger--Heisenberg Quantum Mechanics  in the 1925--1926s,  problems arose on the validity of these postulates in the
dynamical  theory.
A rigorous dynamical description  of these postulates  is still
unknown.
This lack of theoretical clarity hinders the progress in the theory (e.g., in superconductivity and  in nuclear reactions),
and in
numerical simulation of many engineering processes (e.g., of laser radiation and quantum amplifiers) since a computer can solve dynamical equations but cannot  take into account  the postulates.
\smallskip

It is obvious that Postulates I and II
are not compatible with the linear  Schr\"odinger equation. 
We suggest a novel mathematical {\it conjecture on global attractors}
of the {\it coupled (``self-consistent")  nonlinear Maxwell--Schr\"odinger}
equations which clarifies these postulates. 
 This conjecture is confirmed 
for a list of {\it model
Hamilonian nonlinear} PDEs
by
the results obtained since 1990, 
we survey sketchy these results
in Section \ref{sattr} (the details can be found in the survey
 \ci{K2016}). 
However, for the Maxwell--Schr\"odinger equations this conjecture is still an {\it open problem}.
Let us note that these coupled equations
in particular underly modern theory of laser radiation \ci{H1984}--\ci{Kdif2019}.

On the other hand,  the
{\it probabilistic interpretation},
postulated by Born in 1927, agrees satisfactory with the
linear 
Schr\"odinger theory  
in the case of electron beams with high intensity. 
 Newertheless, for very low intensity, the diffraction becomes a random process
 with discrete registration of the diffracted electrons, and  
 the diffraction pattern is obtained by  long-time averaging.
 This 
 was discovered for the first time by 
  Biberman, Sushkin and Fabrikant
\ci {BSF1949} and 
confirmed later by
Chambers, Tonomura, Frabboni,
Bach \& al. \ci { BPLB2013, Chambers60, FGP07, Tonomura89}.
This discrete registration is a {\it genuine nonlinear effect}, and 
we suggest its interpretation  
in the framework 
 of the coupled nonlinear Maxwell--Schr\"odinger
 equations. 
\smallskip

  We discuss these questions and their connections with the theory
 of attractors of Hamilton nonlinear partial differential equations. 
 The {\it nonlinearity is inevitable due to the wave-matter interaction}.
For example,
  the electronic charge density 
generates the corresponding ``own" scalar potential, which 
should be added to the Coulomb potential of molecular nucleus.
Then the Schr\"odinger operator changes with the wave function
that means a nonlinear self-action. 
 \smallskip

We start with a  presentation of the 
Schr\"odinger non-relativistic quantum mechanics and of the Pauli spin theory
for  mathematically oriented readers. This  
is necessary for
the introduction of  nonlinear self-consistent Maxwell--Schr\"odinger equations
and Maxwell--Pauli.
\smallskip

The Schr\"odinger equation is introduced as a wave equation with short-wave solutions propagating
asymptotically along classical trajectories.
The definition of quantum observables is also justified by an asymptotic correspondence with
classic observables.

The electron spin is introduced to describe
the results of the Stern--Gerlach  experiment through
the projective representation of the rotation group $ SO (3) $.
 A simple proof of the covariance of the Pauli equation with respect to the action of the $ SO (3) $ group is given, as well as
relevant conservation laws.

Further we introduce the self-consistent nonlinear 
Maxwell--Schr\"odinger  equations.
We  formulate 
a conjecture which concerns attractors of  
 ``generic" $ G $ -invariant nonlinear Hamiltonian partial differential equations.
We survey the results obtained in 
the 1990--2019 which confirm this conjecture for the model
 equations.
 Then we apply this conjecture to the 
 Maxwell--Schr\"odinger equations for
dynamic interpretation of the basic postulates I – III of quantum mechanics.
 

Finally we calculate
the diffraction amplitude of the electron beam  
in Kirchhoff' approximation. The amplitude  satisfactory agrees with recent experimental data
for the Young two-slit experiment.
The Aharonov--Bohm effect is explained using the Agmon--Jensen--Kato stationary scattering theory.

In  the Appendix we present the “old quantum theory” (1890--1924), which is necessary for
  understanding the emergence of the  Schr\"odinger theory and for the introduction of electron spin.

 
 
 \noindent {\bf Acknowledgements.}
The author thanks Alexander Shnirelman and Herbert Shpon for useful discussions and long-term collaboration.

\section {Planck's law, Einstein's photons and de Broglie 
wave-particle duality} \la {s112}

 The emergence of the Schr\"odinger theory and  the Pauli spin theory was prepared
 by all the
 developments of  ``Old Quantum Mechanics" in the  1890--1925s, which is
 described in the Appendix. A decisive impetus
for the emergence of Schr\"odinger's theory
came from the Planck-Einstein theory of the photons 
and de Broglie conjecture on
wave-particle duality.

\subsection {Planck's law and Einstein's photon} \la {s112}

In 1900, Planck found a fundamental formula for 
spectral intensity of equilibrium radiation
at absolute temperature $ T> 0 $ (Nobel Prize for 1918):
\begin {equation} \la {KP}
I (\omega) = \frac {\hbar \omega ^ 3} {\pi ^ 2c ^ 3} \, \frac {e ^ {- \frac
{\hbar \omega} {kT}}} {1-e ^ {- \frac {\hbar \omega} {kT}}}, 
\end {equation}
where $ \hbar $ is the {\it Planck constant}.
Values of physical constants 
(the electron charge and mass, the Planck constant and the speed of light in the vacuum)
in {\it unrealized Gaussian units}
(Heaviside--Lorentz units (\ci [p. 221] {W2002})) are approximately equal
\be \la {HL}
  e = -4.8 \times 10 ^ {- 10} {\rm esu}, \quad \cm = 9.1 \times 10 ^ {- 28} {\rm g}, \quad
 \hbar = 1.1 \times 10 ^ {- 27} \mbox {\rm erg $ \cdot $ s}, \qquad c = 3.0 \times 10 ^ {10} {\rm cm / s}.
 \ee
In 1905, Einstein proposed a new derivation of the Planck formula
(\ref{KP})
 by a discretisation  with step
\be \la {dE}
\De E = \hbar \om
\ee
for allowed  energies of waves with the frequency $ \om $, see \ci [Section 1.3.4] {K2013}. 
At the same time Einstein proposed to
identify this portion of energy with the energy of a {\it photon} (hypothetical ``particle of light") to explain the {\it photo effect}
(Nobel Prize for 1921). Namely, the light of frequency $ \om $, when falling onto  a metallic surface, it knocks
electrons out of the metal.  Experimental observations give
the maximal kinetic energy of the ``photoelectrons" 
 \be \la {EE}
 W = \hbar \om-A,
\ee
where $ A $ is the {\it work function} \ci [Section 8.4] {K2013}.
Einstein   treated this {\it empirical formula} as the energy balance in the collision of the 
photons with an electron in the metal: 
\begin {equation} \la {Ep}
\textbf {light wave}\quad\psi (x, t) = Ce ^ {i (kx- \om t)}~~ \Lra ~~\textbf {beam of particles (photons) with energy $ E =\hbar\om$}
\end {equation}
The photon energy $\hbar \om$  is {\it partially transmitted} to an  electron of the metal; this electron leaves the metal losing an energy
$A$ to overcome the attraction to the metal.

\subsection {De Broglie wave-particle duality}

 The  energy $ E $ and  momentum $ p $ 
for free {\it non-relativistic} and {\it relativistic} particles
are related by the equations
\be \la {EES}
E = \fr {p ^ 2} {2 \cm}, \qquad \qquad \qquad \fr {E ^ 2} {c ^ 2} = p ^ 2 + \cm ^ 2c ^ 2
\ee
respectively, where $ c $ is the speed of light in the vacuum, and $ \cm $ is the rest mass of particles (for photons $ \cm = 0 $).
  \smallskip
  
  In 1923
 de Broglie suggested the possibility of a wave description of matter
as an {\it antithesis} of the Einstein corpuscular theory of light (photons) (\ref{Ep}).
The main postulate of de Broglie' PhD  was the correspondence
 \begin {equation} \la {dB1}
\textbf {beam of free particles with momentum $ p $ and energy $ E $}
~~ \Lra ~~
 \psi (x, t) = Ce ^ {i (kx- \om t)},  
\end {equation}
where the vector $ (k, \om) $ is some function of the vector $ (p, E) $. The key de Broglie's idea was that
i) this correspondence must be relativistically covariant, and ii) $ kx- \om t $ is a Lorentz-invariant scalar product.
This easily implies that these vectors are proportional: $ (p, E) \sim (k, \om) $ (see \ci [pp 36--37] {K2013}). Finally, the empirical   Planck--Einstein law (\ref{Ep})
gives
 $E=\hbar \om$ which
suggests the ratio
 \begin {equation} \la {dB}
(p, E) = \hbar (k, \om),
\end {equation}
which was experimentally confirmed
by Davisson and Germer in the1924-1927s for  the diffraction of electrons
(see section \ref {sdif} below).

In particular, the de Broglie wavelength is as follows:
\be \la {lam}
\lam = \fr {2 \pi} {| k |} = \fr {2 \pi \hbar} {| p |}.
\ee
This great discovery of de Broglie still plays a key role in nuclear physics in calculating
the energy and momentum of neutrons and other elementary particles
in terms of the wavelength, which is measured in
their diffraction.
\smallskip\\
Now formulas (\ref{EES}) give respectively
\be \la {EES2}
\hbar\om = \fr {\hbar^ 2k^2} {2 \cm}, 
\qquad \qquad \qquad \fr {\hbar^ 2\om^2} {c ^ 2} = \hbar^ 2k^2 + \cm ^ 2c ^ 2.
\ee
 
\section {The Schr\"odinger quantum mechanics} \la {s13}
  The next step was made by Schr\"odinger in the 1925--1926s
by introduction of the Schr\"odinger equation and of stationary orbits.

\subsection{Canonical Quantisation}

{\bf Free particles.}
For the wave function  $\psi (x, t) = Ce ^ {i (kx- \om t)}$ from (\ref {dB1}) it follows that
\be \la {cq0}
i \hbar \pa_t \psi (x, t) = E \psi (x, t), \qquad -i \hbar \na \psi (x, t) = p \psi (x, t).
\ee
Hence, the first
formula (\ref {EES2}) for free non-relativistic particles
together with (\ref{dB}) 
implies the free Schr\"odinger equation
 \be \la {S0}
i \hbar \partial_t \psi (x, t) = - \frac {\hbar ^ 2} {2 \cm}
\De \psi (x, t).
\ee
Similarly, the second formula (\ref {EES2})
 for relativistic particles
 together with (\ref{dB}) 
  implies
 the free Klein--Gordon equation 
 \be \la {KG0}
\frac
1 {c ^ 2} [i \hbar \pa_t] ^ 2 \psi (x, t) = [(- i \hbar \na) ^ 2 +
\cm ^ 2c ^ 2] \psi (x, t).
\ee
Mnemonically,
 equations (\ref {S0}) and (\ref {KG0}) are derived from expressions for energies (\ref {EES})
by replacements
\be \la {cq}
E \mapsto \hat E: = i \hbar \pa_t, \qquad p \mapsto \hat p: = - i \hbar \na,
\ee
which are called the {\it Canonical Quantization}.
\smallskip\\
{\bf Bound particles.}
Now let us consider particles 
in the {\it external Maxwell field} 
\be \la {MF}
 \bE^\ext (x, t) = - \fr 1c \dot \bA^\ext (x, t) - \na A^\ext_0 (x, t), \, \qquad \bB^\ext ( x, t) = \rot \5 \5 \bA^\ext (x, t)
  \ee
with magnetic potential $ \bA^\ext (x, t) = (\bA^\ext_1 (x, t), \bA^\ext_2 (x, t), \bA^\ext_3 (x, t)) $ and scalar potential $ A^\ext_0 (x, t) $.
\smallskip\\
For nonrelativistic particles,
the energy (Hamiltonian) is expressed by the formula \ci [(12.90)] {K2013}
\be \la {ES}
E = \cH (x, p, t) = \fr 1 {2 \cm} [p- \ds \frac ec \bA^\ext (x, t)] ^ 2 + eA^\ext_0 (x, t ),
\ee
where $ e <0 $ is the charge of an electron.
In this case, the canonical quantization (\ref {cq}) leads to the Schr\"odinger equation
\be \la {S}
i \hbar \dot \psi (x, t) =
H (t) \psi (t): =
\fr 1 {2 \cm} [- i \hbar \na- \ds \frac ec \bA^\ext (x, t)] ^ 2 \psi (x, t) + eA^\ext_0 (x, t ) \psi (x, t).
\ee
For relativistic particles the energy is expressed by the formula \ci [(12.93)] {K2013},
\be \la {EKG}
[\frac Ec- \ds \frac ecA^\ext_0 (x, t)] ^ 2 = [p- \ds \frac ec
\bA^\ext (x, t)] ^ 2+ \cm ^ 2c ^ 2,
\ee
which
leads to the Klein--Gordon equation
\be \la {KG}
[\frac {i \hbar \pa_t} c- \ds \frac ecA^\ext_0 (x, t)] ^ 2 \psi (x, t) = [- i \hbar \na- \ds \frac ec
\bA^\ext (x, t)] ^ 2 \psi (x, t) + \cm ^ 2c ^ 2 \psi (x, t).
\ee

\subsection{Wentzel--Kramers--Brillouin quasiclassical asymptotics}
The derivation of   equations (\ref {S}) and (\ref {KG})
 relies 
only on  de Broglie's relativistic arguments and formulas (\ref {EES}) of classical electrodynamics.
However,  a deeper basis for these equations lies in  short-wave asymptotics.
For example, the Hamiltonian equations corresponding to the nonrelativistic Hamiltonian (\ref {ES}) are equivalent to the Lorentz equation
(\ref {L}) (see \ci [Section 12.6] {K2013}), where the 
Maxwell fields are expressed in terms of potentials by formulas (\ref{MF}).
The key argument for the Schr\"odinger equation (\ref {S}) was the short-wave 
Wentzel--Kramers--Brillouin asymptotics (1926)
for solutions to the Cauchy problem
with fast-oscillating initial data $ \psi (x, 0) = a_0 (x) e ^ {- iS_0 (x) / \hbar}$:
\be \la {WKB}
{\rm (WKB)} \qquad \qquad \qquad \qquad \qquad \psi (x, t) \sim a (x, t) e ^ {- iS (x, t) / \hbar}, \qquad \hbar \ll 1. \qquad \qquad \qquad \qquad
\ee
The main fact is that if $ a_0 (x) = 0 $ outside the ball $ | x-x_0 | \le \ve $ with small $ \ve> 0 $, then
\be \la {tt}
\! \! \! \! \!
\left |
\ba {l}
\mbox {\bf
$ a (x, t) = 0 $ outside a thin tubular neighborhood of the trajectory of $ x (t) $}
\medskip \\
\mbox {\bf of the Lorentz  equation (\ref {L}) with initial data}
\medskip \\
\qquad \qquad x (0) = x_0, \qquad p (0): = \cm \dot x (0) + \fr ec A (x (0), 0) = a_0 (x_0) \na S_0 (x_0 )
\ea \right |,
\ee
see
 \ci [Sections 3.2 and 13.2] {K2013}.
In other words, short-wave solutions to the Schr\"odinger equation
(\ref {S}) {\bf propagate along the trajectories of the non-relativistic classical Lorentz equation} (\ref {L}).
This is due to the fact
that
  \smallskip
  

  a) The phase function $ S (x, t) $ satisfies the Hamilton--Jacobi equation
  \be \la {HJ}
  \pa_t S (x, t) = \cH (x, -i \na S (x, t), t),
  \ee
  where $ \cH (x, p, t) $ is the Hamiltonian (\ref {ES}), and
   \smallskip

  b)
Solutions of the Hamilton-Jacobi equation are obtained
by integrating along the trajectories of the corresponding Hamiltonian system \ci [Section 46] {A1989}, 
which   coincides in this case with the Lorentz equation (\ref {L}), see \ci[Section 12.6]{K2013}.
   \smallskip \\
The Hamilton--Jacobi equation (\ref {HJ}) is obtained by substituting the asymptotics (\ref {WKB}) into the Schr\"odinger equation (\ref {S}) 
and setting $ \hbar = 0 $.
   
  
\brs
 {\rm
 i) The structure
of the Hamilton-Jacobi equation (\ref {HJ}) 
dictates uniquely the rules of canonical quantization
 (\ref {cq}).
\smallskip \\
ii) Asymptotics (\ref {WKB})  elucidate a new light into Thompson's experiments with electrons
in the 1893--1897s. It appears that he observed exactly short-wave
asymptotics of cathode rays, which makes the concept of elementary particles in quantum theory  problematic.
\smallskip \\
iii)
 These {\it quasiclassical asymptotics} are examples of the Bohr{\it correspondence principle} \ci [Section 3.3.3] {K2013}.
 We will call solutions of the type (\ref {WKB}), (\ref {tt}) {\it quasiclassical}.
 Similar asymptotics hold for solutions of reativistic
 equations: for  the Klein--Gordon equations and the Dirac equations (Pauli 1932), and
 for general hyperbolic systems
  (Lax 1957, Maslov 1965, H\"ormander, and others).
}
\ers
\section {Quantum Observables} \la {s14}

Quantum observables are invariants of the 
Schr\"odinger dynamics. The invariance is caused by the symmetry of external potentials:
the electron energy is conserved if the external potentials are independent of time,
the charge is conserved at any real potentials, the projection of the impulse on a certain direction is preserved if the potentials are
invariant with respect to shifts in this direction,
and the projection of angular momentum onto a certain direction is preserved if the potentials are invariant (in a certain sense) on
rotations
around this direction.

The {\it Correspondence Principle} is that
the quantum and corresponding classical observables asymptotically coincide as $ \hbar \to 0 $,
see (\ref{pL2}).

\subsection {Hamiltonian structure and energy}
The linear Schr\"odinger equation (\ref {S}) can be written in the Hamiltonian form as
\be \la {HS}
i \hbar \dot \psi (t) = \fr 12 D_ \psi \cH (\psi (t), t) = H (t) \psi (t), \qquad t \in \R,
\ee
where $ D_ \psi $ is a variational derivative, and the Hamilton functional $\cH$ (= quantum energy $\ov E (t)$) 
reads as
\beqn \la {HSH}
\ov E (t) &: = & \cH (\psi (t), t): = \langle \psi (t), H (t) \psi (t) \rangle
\nonumber \\
\nonumber \\
& = & \int \Big[\fr 1 {2 \cm} | [-i \hbar \na- \fr ec \bA^\ext (x, t)] \psi (x, t) | ^ 2 + eA^\ext_0 (x, t) | \psi (x, t) | ^ 2\Big] dx.
\eeqn
Therefore, the energy is conserved if the external potentials are independent of time.

\subsection {Charges and currents}
 A comparison (\ref {HSH}) with energy in electrostatics shows that the  
electric charge density should be defined as
\be \la {rho}
\rho (x, t) = e | \psi (x, t) | ^ 2.
\ee
Now equation (\ref {HS}) implies the conservation of the total charge
\be \la {qcc}
Q (t): = \int \rho (x, t) dx \equiv \const.
\ee
This follows from the symmetry of the Schr\"odinger operator $ H (t) $:
\be \la {cce2}
\dot Q (t): = e \langle \dot \psi (t), \psi (t) \rangle + e \langle \psi (t), \dot \psi (t) \rangle =
- \fr e {i \hbar} \langle H (t) \psi (t), \psi (t) \rangle + \fr e {i \hbar} \langle \psi (t), H (t) \psi (t) \rangle \equiv 0,
\ee
where
the brackets $ \langle \cdot, \cdot \rangle $ mean anti-Hermitian scalar product.

For one electron $ \ds \int \! \! \rho (x, t) dx \! = \! e $, whence according to (\ref {rho}) the {\it normalization condition} holds
\be \la {nor}
\int | \psi (x, t) | ^ 2dx = 1.
\ee
Further, the current density should be defined as \ci [(21.12)] {FL}):
\be \la {j}
\bj (x, t) = \fr e \cm \rRe \{\ov \psi (x, t) [- i \hbar \na- \fr ec \bA^\ext (x, t)] \psi (x, t) \},
\ee
since this density together with (\ref {rho}) satisfies {\it the continuity equation}
\be \la {cce}
\dot \rho (x, t) + \dv \bj (x, t) \equiv 0.
\ee
This can be either verified  by direct differentiation \ci [Section 3.4] {K2013} or derived from
Noether's general theorem on invariants \ci [Section 13.4.3] {K2013}
using the $ U (1) $-invariance of
Hamiltonian (\ref {HSH}) with respect to the action
$ \psi (x) \mapsto e ^ {i \theta} \psi (x) $, where $ \theta \in (0,2 \pi). $

\subsection {Quantum momentum  and angular momentum}

The quantum momentum and the angular momentum in Schr\"odinger's theory  are defined 
for any state $ \psi (t) $ as {\it mean values}
\be \la {pL}
\ov p (t): = \langle \psi (t), \hat p \psi (t) \rangle, \qquad \ov L (t): = \langle \psi (t), \hat L \psi (t) \rangle,
\ee
where $ \hat p := -i \hbar \na $, $ \hat L: = \hat x \wedge \hat p $ are
{\it self-adjoint operators} of the momentum and the angular momentum, $ \hat x $ is the multiplication operator by $ x $.
 
 \subsection {Correspondence Principle}
 The name of the {\it momentum, angular momentum}, as well as
 {\it energy} (\ref {HSH}), are
 justified by the {\it Correspondence Principle}: for the quasiclassical solutions (\ref {WKB})
 with small $ \hbar \ll 1 $ and $ \ve \ll 1 $,
 \be \la {pL2}
 \! \! \! \!
\ov E (t) \! \sim \! \fr 1 {2 \cm} [p (t) - \fr ec \bA^\ext (x (t), t)] ^ 2 + eA^\ext_0 (x (t), t), \quad
\ov p (t) \! \sim \! \cm \dot x (t) + \fr ec \bA^\ext (x (t), t), \quad \ov L (t) \! \sim \! x (t) \wedge p (t),
\ee
 see \ci [Section 3.3.3] {K2016}.
  
  
  \subsection {Conservation Laws}
The formal mathematical motivation for
these definitions of observables for the Hamiltonian system (\ref {HS})
consists in  {\it conservation laws} \ci [Section 3.3.2] {K2016}: 
for solutions to the Schr\"odinger equation (\ref {S}) 
\smallskip\\
i) the energy
$ \ov E (t) = \const $ if the Maxwell potentials are independent of time, 
\smallskip\\
ii) the ``mean momentum"
$ \ov p_n (t) = \const $ if the potentials are independent of $ x_n $, and 
\smallskip\\
iii) the ``mean angular momentum"
  \be \la {Lc2}
 \ov L_n (t) = \const
 \ee
if the potentials are invariant with respect to rotations around the $ x_n $ axis; i.e.,
 \be \la {Rn}
 \bA^\ext (R_n (\vp) x) \equiv R_n (\vp) \bA^\ext (x), \quad A^\ext_0 (R_n (\vp) x) \equiv A^\ext_0 ( x), \qquad \vp \in [0,2 \pi],
 \ee
 where $ R_n (\vp) $ is the rotation of the space $ \R ^ 3 $ through the angle $ \vp $ around the axis $ x_n $, and the direction of rotation is determined by the gimlet rule.
 \smallskip \\
{\bf Example.} For a uniform magnetic field $ B = (0,0, B_3) $
 the vector potential
has the form 
\be \la {Aext}
\bA^\ext (x) = \fr 12B_3 (-x_2, x_1,0)
\ee
and satisfies the condition (\ref {Rn}) with $ n = 3 $.
 \smallskip
\subsection{Proof of conservation laws}
The energy conservation was proved above. 
The  conservation of momentum and angular momentum follows from  {\it commutation relations}.
Namely, if the potentials are independent of $ x_n $, 
then the commutator
  $ [\hat p_n, H (t)] = 0 $ and differentiation yields 
  \be \la {dp}
  \dot {\ov p} _n (t) = \langle \dot \psi (t), \hat p_n \psi (t) \rangle + \langle \psi (t), \hat p_n \dot \psi (t) \rangle = \fr 1 {i \hbar} \langle \psi (t), [\hat p_n, H (t)] \psi (t) \rangle = 0,
  \ee
where we used the Schr\"odinger equation (\ref {HS}).
Similarly,  (\ref {Rn}) implies the commutation 
 \be \la {LnH}
 [\hat L_n, H (t)] = 0,
 \ee
because $ \hat L_n = -i \hbar \pa_ \vp $ in cylindrical coordinates with the $ x_n $ axis,
while $ H (t) $ is a differential operator
with coefficients independent of $ \vp $ ({\bf Exercice}).
Therefore,
  \be \la {dp2}
  \dot {\ov L} _n (t) = \langle \dot \psi (t), \hat L_n \psi (t) \rangle + \langle \psi (t), \hat L_n \dot \psi (t) \rangle = \fr 1 {i \hbar} \langle \psi (t), [\hat L_n, H (t)] \psi (t) \rangle = 0.
  \ee
{\bf Definition.} A quantum observable is {\it a  quadratic form}
  \be \la {Kp}
  K (\psi) = \langle \psi, K \psi \rangle,
  \ee
where $ K $ is some self-adjoint operator in $ L ^ 2 (\R ^ 3) $.
  \medskip \\
As above, the observable $ K (\psi (t)) $ is conserved for solutions of the Schr\"odinger equation (\ref {HS}) if
  \be \la {Kpc}
  [K, H (t)] = 0, \qquad t \in \R.
  \ee
  

\section{Bohr's postulates, stationary orbits  and attractors}
   In 1913, Bohr formulated two fundamental postulates of quantum theory of atoms:
\smallskip \\
{\bf I.} An atom is always in one of quantum stationary orbits, and
sometimes it jumps from one stationary orbit to another:
in the Dirac notation 
\be \la {B1}
| E_n \rangle \mapsto | E_ {n '} \rangle.
\ee
{\bf II.} The atom does not radiate in  stationary orbits. Every  jump
is followed by a radiation of an electromagnetic wave with the
frequency 
\be \la {B21}
 \om_ {nn '} = \fr {E_ {n'} - E_n} \hbar = \om_ {n '} - \om_n, \qquad \om_n: = E_n / \hbar,
\ee

With the discovery of the Schr\"odinger theory in 1926,
the question arose about the validility of these Bohr's axioms in the new theory.
   
\subsection{Schr\"odinger theory of stationary orbits}
Besides the equation for the wave function, the Schr\"odinger theory contains quite a
nontrivial definition of stationary orbits in the case when
 {\bf the Maxwell external potentials do not depend on time:} 
\be\la{stext}
\bA^\ext (x,t)\equiv \bA^\ext (x),\qquad A^0(x,t)\equiv A^0(x).
\ee
In this case $H(t)\equiv H$.
\bd
{\bf Stationary orbits} are solutions  of the form
\be \la {SSO}
\psi (x, t) \equiv \vp (x) e ^ {- i \om t}, \qquad \om \in \R
\ee
to the Schr\"odinder equation (\ref{S}).
\ed

Substitution 
into the Schr\"odinger equation (\ref{Sc0}) leads to the famous {\bf eigenvalue problem}
\be\la{eipr}
H\vp=\hbar\om\vp.
\ee
Definition of stationary orbits (\ref{SSO}) is
 rather natural, since then $ | \psi (x, t) | $ does not depend on time. This definition 
 probably
 was suggested
by the de Broglie wave function  for{\it free particles}
$ \psi (x, t) = Ce ^ {i (kx- \om t) } $, which factorizes as $ Ce ^ {ikx} e ^ {- i \om t} $. Namely, in the case of  {\it bound particles}
it is natural to change the spatial factor $ Ce ^ {ikx} $, since the spatial properties have changed and ceased to be homogeneous.
On the other hand, the homogeneous time factor $ e ^ {- i \om t} $ must be preserved, since the external potentials are independent of time.
However, these ``algebraic arguments" do not 
withdraw the question  on agreement of the Schr\"odinger definition
with the  Bohr postulate (\ref {B1})!
\smallskip

Thus, a {\bf problem} arises on the mathematical interpretation of the Bohr postulate (\ref {B1}) in the Schr\"odinger theory. 
   One of the {\bf simplest interpretation} of the jump (\ref{B1})
is the  {\it long-time asymptotics} 
\be \la {BS}
\psi (x, t) \sim \psi_ \pm (x) e ^ {- i \om_ \pm t}, \qquad t \to \pm \infty,
\ee
{\it for each finite energy solution},
 where
 $ \om _- = \om_n $ and $ \om _ + = \om_ {n '} $. However, 
 {\bf for  linear Schr\"odinger equation} (\ref{Sc0})
such asymptotics are obviously  wrong  due to the {\it superposition principle}:
for example, for solutions of the form $ \psi (x, t) \equiv \psi_1 (x) e ^ {- i \om_1 t} + \psi_2 (x) e ^ {- i \om_2 t} $ with $\om_1\ne\om_2$.
It is exactly this contradiction which shows that the linear 
Schr\"odinger equation alone cannot  serve as a basis for the 
theory compatible
with the Bohr postulates.

{\it Our main conjecture} is that these postulates are inherent properties of the nonlinear Maxwell--Schr\"odinger equations, see (\ref{MS}) below. This conjecture
is suggested by the following perturbative arguments.

\subsection{Bohr' postulates and asymptotics (\ref{BS}) by perturbation theory}   
   
The remarkable success of the Schr\"odinger theory was the explanation
of the {\it Bohr' postulates and asymptotics}
(\ref{BS})
 by perturbation theory
in the case of {\it static external potentials} (\ref{stext}).
Namely, in this case  the Schr\"odinger equation (\ref {S}) reads
 \be \la {Sc0}
 i \hbar \dot \psi (x, t) =H\psi (x, t).
\ee
For 
``sufficiently good"
external potentials and initial conditions,
any finite energy solution
can be expanded in  eigenfunctions
 \be \la {Sexp}
 \psi (x, t) = \sum_n C_n \psi_n (x) e ^ {- i \om_n t} + \psi_c (x, t), \qquad
 \psi_c (x, t) =
 \int C (\om) e ^ {- i \om t} d \om,
\ee
where the integration is performed over the continuous spectrum of the Schr\"odinger operator $ H $, and 
\be\la{psic}
\psi_c (x, t)\to 0,\quad t\to\pm\infty,\qquad x\in\R^3,
\ee
see, for example, \ci [Theorem 21.1] {KK2012}.
The
substitution of this expansion into the expression for currents (\ref {j}) gives 
 \be \la {jexp}
\bj (x, t) = \sum_ {nn '} \bj_ {nn'} (x) e ^ {- i \om_ {nn '} t} + c.c. + \bj_c (x, t),
\ee
where $ \bj_c (x, t) $ contains the continuous frequency spectrum.
These currents enter into the
Maxwell equations 
in the entire space of $ \R ^ 3 $:
 in
the Heaviside--Lorentz units (\ci [p. 781] {Jackson}) 
\be \la {mhl-1}
\dv \bE (x, t) = \rho (x, t), \, \, \, \rot \bE (x, t) = - \fr 1c \dot \bB (x, t), \, \, \dv \bB (x, t) = 0, \, \, \rot \bB (x, t) = \fr 1c [\bj (x, t) + \dot \bE (x, t)].
\ee
The second and third equations imply the Maxwell representations
$ \bB (x, t) = \rot \5 \5 \bA (x, t)$ and 
$ \bE (x, t) = - \fr 1c \dot \bA (x, t) -\na A ^ 0 (x, t). $
Then
in the Coulomb gauge $ \dv \bA (x, t) \equiv 0 $,
the Maxwell equations (\ref {mhl-1}) are equivalent to the system
\be \la {Ma-1}
\fr 1 {c ^ 2} \ddot \bA (x, t) = \De \bA (x, t) +\fr1c P\bj (x, t), 
\quad \De A^0(x,t)= -\rho (x, t),\qquad x \in \R ^ 3,
 \ee
where  $P$ is the {\it orthogonal projection}   
in the Hilbert space $L^2(\R^3)\otimes\R^3$ onto free-divergent vector fields.
\smallskip

Thus, the currents (\ref{jexp}) on the right  of 
the
Maxwell equations (\ref{Ma-1}) contain,
besides the continuous spectrum, only
{\bf discrete frequencies} $ \om_ {nn '} $.
Hence, the {\bf discrete spectrum} of the corresponding 
Maxwell radiation field $\bA(x,t)$ also contains only these frequencies $ \om_ {nn '} $.
This proves the Bohr rule (\ref {B21})
{\it in the first order of perturbation theory}, since this calculation ignores the back reaction  of radiation onto the atom. 
\smallskip

Moreover, these arguments also justify  
{\it the jumps  (\ref {B1}) as
the long-time asymptotics (\ref {BS})}. Namely, 
the currents (\ref{jexp}) on the right  of 
the Maxwell equation from (\ref {MS}) produce the radiation when 
nonzero frequencies  $ \om_ {nn '} $ are present. However, this radiation
cannot last forever since the total energy is finite. Hence, in the long-time 
limit  only $ \om_ {nn '}=0 $ should remain, which means exactly {\it one-frequency asymptotics} (\ref{BS}) 
by (\ref{psic}).

\section {Stationary orbits of  electron in atom}
In the case of one electron in an atom
with nucleous charge   $ eZ $ 
the
electrostatic Coulomb potential of the nucleus  reads as $ A^0 (x) = - \fr {Ze } {| x |} $,
while the magnetic potential can be neglected. Now the Schr\"odinger equation (\ref {HS}) becomes
\be \la {HS2}
i \hbar \dot \psi (t) = H \psi (t), \qquad H = - \fr {\hbar ^ 2} {2m} \De- \fr {e ^ 2Z} {| x |}.
\ee
{\it Stationary orbits} of such an electron {\it are defined} as solutions of the form 
(\ref{SSO}).
Substitution into equation (\ref {HS2}) results in
the eigenvalue problem
\be \la {evp}
H \psi (x) = \hbar \om \psi (x).
\ee
 The ``mean energy" of a stationary orbit is equal to the eigenvalue due to normalization (\ref {nor}),
 \be \la {En}
\ov E: = \cH_S (\psi): = \langle \psi (t), H \psi (t) \rangle = \hbar \om \langle \psi (t), \psi (t) \rangle.
 \ee
 These eigenvalues were first calculated by Schr\"odinger
 \ci [I] {Schr1926}
 using separation of variables in spherical coordinates.
Later, these calculations were simplified using
irreducible
representations for the Lie algebra of the rotation group $ SO (3) $
to find the spectrum of the spherical Laplacian \ci [Chapter 6] {K2013}.
 The result is the Balmer formula.
 \be \la {Bal}
\om_n = - \fr {b} {n ^ 2}, \quad n = 1,2, ..., \quad b = 2 \pi cR \approx 2 \cdot 10 ^ {16} {\rm s} ^ {- 1}
\ee
 with the Rydberg constant
  \be \la {RID2}
R = \fr {me ^ 4Z ^ 2} {4 \pi \hbar ^ 3c}, \qquad \mbox {(the current value is $ R = 109737.31568527 \, \, {\rm cm} ^ {- 1} $)} ,
 \ee
which completely coincides with the formula of ``Old Quantum Mechanics" (\ref {RID}).
This confirmed both the Schr\"odinger theory,
and the ``Old Quantum Mechanics".
The eigenfunctions and eigenvalues are numbered with the same quantum numbers $ n, l, m $ as
in  formulas (\ref {ELr4}), (\ref {Lm}).
However now the
{\it value $ l = n $ for the angular momentum is excluded}, and the formulas (\ref {ELr4}),
 (\ref {ELr5}) are slightly modified
(see, e.g., \ci [Ch. 6] {K2013}):
 \beqn \la {nlmE}
 \left \{
 \ba {l}
 H \psi_ {nlm} = E_n \psi_ {nlm}, \quad E_n = - \ds \fr {\cm e ^ 4} {2 \hbar ^ 2n ^ 2}
 \\
 \\
 \hat L ^ 2 \psi_ {nlm} = \hbar ^ 2 l (l + 1) \psi_ {nlm}, \, \,
\hat L_3 \psi_ {nlm} = \hbar m \psi_ {nlm} \\
  \ea \right | \, \, \,
n \! = \! 1,2, \dots; \, \, l \! = \! 0,1, \dots, n \! - \! 1; \, \, m \! = \! - l , \dots, l.
 \eeqn
Here $ \hat L ^ 2: = \hat L_1 ^ 2 + \hat L_2 ^ 2 + \hat L_3 ^ 2 $.
The indices $ n, l, m $ are called, respectively, the {\it main, azimuth} and {\it magnetic}
 quantum numbers.
Let us note that the  operators
$ H $, $ \hat L ^ 2 $, $ \hat L_3 $ have common eigenfunctions due to their commutation
\be \la {HLLcom}
[H, \hat L ^ 2] = [H, \hat L_3] = [\hat L ^ 2, \hat L_3] = 0
\ee
which holds by the spherical symmetry of the Coulomb potential of the nucleus.
These commutations imply  that the observables $ \hat L ^ 2 (\psi) $ and $ \hat L_3 (\psi) $ are conserved along the solutions to the Schr\"odinger equation (\ref {HS2}).

\section {Atom in magnetic field} \la {s7S}
In 1895, Zeeman observed a splitting of the spectral lines (\ref{B21}) in a magnetic field.

\subsection{Normal Zeeman effect}
The ``normal Zeeman effect" is the splitting of one line $\om_{nn'}$
into three separate lines
\be \la {Zeemantr}
\om_{nn'}- d\om_L,\qquad d= 0, \pm 1,
\ee
where 
\be\la{Larf}
\om_L=\ds\fr {e B_3} {2 \cm c},
\ee
is the {\it Larmor frequency}.
This normal effect
was successfully explained in 
the classical Lorentz model 
(Section \ref {s11}) as well as in the ``Old Quantum Theory" (Section \ref {s125}).

The Schr\"odinger theory leads to the same result.
Namely,
for an atom in a magnetic field with a potential $ \bA^\ext (x) $, the Schr\"odinger equation (\ref {HS}) 
reads
\be \la {HA}
i \dot \psi (t) = H \psi (t), \qquad
H = \fr 1 {2 \cm} [- i \hbar \na- \fr ec \bA^\ext (x)] ^ 2- \fr {e ^ 2Z} {| x |}.
\ee
In particular, for a uniform external magnetic field $ B = (0,0, B_3) $, the  vector potential
has the form (\ref {Aext}),
and the Schr\"odinger equation (\ref {HA})
becomes
\be \la {HAB}
i \dot \psi (t) = H_B \psi (t), \qquad
H_B = - \fr {\hbar ^ 2} {2 \cm} \De- \fr {e ^ 2Z} {| x |} - \fr e {2 \cm c} B_3 \hat L_3 = H- \om_L\hat L_3,
\ee
when the quadratic  terms with $ \fr {e ^ 2} {c ^ 2} B ^ 2 $ are neglected.
\smallskip

Now  (\ref {nlmE}) implies that
\be \la {nlmE2}
\left \{
\ba {l}
 H_B \psi_ {nlm} = E_ {nm} \psi_ {nlm}, \, \, E_ {nm} =\hbar\om_{n}  - \om_L\hbar m
 \\
 \\
 \hat L ^ 2 \psi_ {nlm} = \hbar ^ 2 l (l + 1) \psi_ {nlm}, \, \,
\hat L_3 \psi_ {nlm} = \hbar m \psi_ {nlm}
 \ea \right | \,
n \! = \! 1,2, \dots; \, l \! = \! 0,1, \dots, n \! - \! 1; \, m \! = \! l, \dots, l.
 \ee
which coincides with (\ref {ELr42}) except for the range of  $l$.
Thus, the spectral lines (\ref {B21}) are shifted by  $- \om_L (m'-m)$ and 
are given by 
\be \la {See}
\om_{nn'}- \om_L (m'-m).
\ee
These spectral lines coincide with  the result (\ref{See2})  in  “Old Quantum Mechanics”.

\subsection{The selection rules}
Let us show that the selection rules (\ref {prot2}) also hold
in the Schr\"odinger theory
in the sense that
the intensity of  other lines vanish. This follows from
formulas
for the intensity of  spectral lines \ci [(45.13), (45.21)] {Schiff1955},
\ci [(7.48), (7.52)] {K2013}. 
Namely, the intensity of 
the line with frequency $ \om_ {n'm '} - \om_ {nm} = (E_ {n'm'} - E_ {nm}) / \hbar $
is proportional to the sum
\be \la {dimos}
   I_{nmn'm'}:= \sum_{l=0}^{n-1} \sum_{l'=0}^{n'-1}|d_ {nlmn'l'm'}|^2
\ee
where  $d_ {nlmn'l'm'}$ is the  {\it dipole moment}
\be \la {dimo}
    d_ {nlmn'l'm'} = \langle \psi_ {nlm}, x \psi_ {n'l'm'} \rangle = \int \psi_ {nlm} x \ov \psi_ {n'l'm'} dx
\ee
 In the spherical coordinates $ r, \vp, \theta $ with the axis $ x_3 $,
all eigenfunctions of the operator $ \hat L_3 = -i \hbar \pa _ {\vp} $
have the form $ c (r, \theta) e ^ {im \vp} $
with integer $ m $
and with the corresponding eigenvalues $ \hbar m $. Hence, 
with a suitable choice of the multiplier, the eigenfunctions $\psi_ {nlm}$ read
\be\la{psinlm}
\psi_ {nlm} = c_ {nlm} (r, \theta) e ^ {im \vp}
\ee
and the corresponding eigenvalue equals $\hbar m$.
Hence,
in the spherical coordinates the
internal integral 
over $ \vp \in [0,2 \pi] $ in (\ref{dimo})
equals zero 
for $ | m'-m |> 1 $, since coordinates of the vector
$ x=(x_1,x_2,x_3) $ contain harmonics $ e ^ {ik \vp} $ only
with $ k = 0, \pm 1 $. 
\smallskip

Thus, the normal Zeeman effect in the Schr\"odinger theory 
can be completely explained.
However, the anomalous Zeeman effect is not explained by this theory.
This problem was solved only after the introduction
in the Schr\"odinger equation 
of the {\it electron spin} and of an additional term
(``Russell-Saunders coupling"), describing the spin interaction
 with the angular momentum $ \hat L $, see \ci [Section 10.3] {K2013}.
  
\br
{\rm
The Schr\"odinger operator (\ref {HAB}) is generalized in a natural way
to the case of  a slowly varying magnetic field $ B (x, t) $,
  \be \la {HAB2}
H_B = - \fr {\hbar ^ 2} {2 \cm} \De- \fr {e ^ 2Z} {| x |} - \fr e {2 \cm c} B (x (t), t) \cdot \hat L,
\ee
where $ x (t) $ is the trajectory of an atom in space.
 }
 \er

  \section {Own magnetic moment of electron} \la {s15}
The introduction of  
electron spin  was
suggested by 
experiments of Einstein--de Haas, Stern--Gerlach,
and Lande's vector model,
and
Bohr theory of the  Mendeleev periodic table of chemical elements.

\subsection {Einstein-de Haas experiments}
In 1915,
the presence of the intrinsic magnetic moment of an electron was demonstrated by
experiments of Einstein--de Haas, who measured the
ratio of the magnetic moment of the electrons of an atom to their angular (“mechanical”) moment.
The result, very surprisingly, 
   did not fit the classical theory. The measurements were based on the observation of torsional vibrations of a ferromagnetic
rod suspended on a thin filament inside the solenoid.
\smallskip \\
The classical angular and magnetic moments of the electrons of an atom are determined similarly to (\ref {LM1}),
\be \la {LM}
L: = \cm \sum_k x_k \wedge v_k, \qquad M: = \fr e {2c} \sum_k x_k \wedge v_k,
\ee
 where $ x_k $ are the positions of electrons of the atom, and $ v_k $ are their velocities. therefore
 \be \la {MLe2}
 M = \fr {e} {2 \cm c} L,
\ee
similarly (\ref {MLe}).
In stationary states of the atom one has
$ \ov {\sum_k v_k} = 0 $, where the bar  means the {\it average over time}. Therefore, formulas like (\ref {LM}) are suitable
for
definition of {\it time-averaged} values
$ \ov M $ and $ \ov L $ regardless of the choice of the origin and position of  atom.
Hence, the same is true for 
{\it time-averaged } values of the angular and magnetic moments
$ \ov L_s $ and $ \ov M_s $
{\it of all electrons} of a macroscopic body,
and
so the summation is now carried over all electrons of the rod. Thus,
\be \la {ML2}
\ov M_s = \fr {e} {2 \cm c} \ov L_s.
\ee
When a current is turned on in the solenoid, the magnetic field orients the magnetic moments of all atoms of the rod,
so that the total magnetic moment of atoms instantly becomes maximal, and accordingly, the total angular momentum of electrons
also instantaneously increases, since the magnetic moment of the nuclei in these experiments is negligible.
But then
 the conservation of the total angular momentum of the rod (electrons together with the crystal lattice)
leads to the opposite rotation of the lattice, to which the suspension thread is attached.
Changing the frequency of the alternating current in the solenoid
one can determine the resonant frequency by the maximal amplitude of the rotational oscillations of the rod,
which allows us to find
the ratio $ \ov M / \ov L $, which enters the equation of oscillations \ci {LPB}. The result was 
in contradiction with (\ref {ML2}):
\be \la {MLg}
\ov M_s = g \fr {e} {2 \cm c} \ov L_s, \qquad g> 1.
\ee
In more accurate experiments (Beck, Arvidson, Klaasen) it turned out 
that the
{\it gyroscopic ratio} $ g \approx 2 $.
This had led to the assumption that the electron itself has angular and magnetic moments with this ratio.

\subsection {Vector model of Lande}
In 1921
Lande suggested a {\it phenomenological classical model} for the
description of the {\it anomalous Zeeman effect} in the framework of ``Old Quantum Mechanics",
treating
the electron's own magnetic moment as the electron rotation around an axis
\ci [Chapter VI, Section 2] {Born1951},
\ci [Section 14.7] {K2013}.
Lande's formula produced a result that coincides remarkably well with the 
experimental observations of the  
anomalous Zeeman effect.

\subsection {Stern--Gerlach experiment and Goldsmith--Uhlenbeck hypothesis}

In 1922,  Stern and Gerlach conducted a
crucial and clarifying
experiment on the splitting of a beam of silver atoms
in {\it two parts}
in a {\it strongly inhomogeneous magnetic field}.
Formula (\ref {MH22})
implies that
an inhomogeneous magnetic field $ B (x, t) $
acts 
on an atom with magnetic moment
$ M $ with the force
\be \la {FM} 
F (t) \approx M \na B (x (t), t),
\ee
and the Lorentz force is $ \fr ec \dot x (t) \wedge B (x (t), t) = 0 $ due to the neutrality of the atom.
Hence,
the splitting of atomic beam
means that
electrons in  atoms of these two beams
were in states with  different magnetic moments,
since the magnetic moment of the nucleus is negligible ($ \sim M / 1800 $).
Later, similar experiments were conducted with hydrogen atoms.

In 1925, Goldsmith and Uhlenbeck suggested an explanation for this fact by an ``own magnetic moment" of electron
which 
can be only  in two states:
the magnetic moment is parallel to the magnetic field
in one state, and  antiparallel in the other.

 Particularly puzzling was the
splitting exactly in two beams because,
according to the “Old Quantum Mechanics”, the splitting is possible only into an odd number of components.
Namely, according to (\ref {Lm}) the
angular momentum of an electron in an atom takes
discrete values
$ \hbar l $ with integer $ l $, and the magnetic moment at a fixed $ l $ takes
$ 2l + 1 $ different
values of $ \fr {e \hbar} {2 \cm c} m $, where $ m = -l, ..., l $ with  step  $ 1 $,
as in the later-developed Schr\"odinger theory.

Respectively, Goldsmith and Uhlenbeck interpreted the
splitting in two levels in the Stern-Gerlach experience
as the presence of 
electron states with $ l = 1/2 $
and with the corresponding angular momentum values $\pm \hbar / 2 $.
At the same time,
they
 postulated the electron ratio (\ref {MLg}) with the experimentally measured value  $ g = 2 $:
\begin {list} {} {\itemsep = 0pt \topsep = 2pt \parsep = 0pt}
\item []
\qquad\qquad{\it
The electron has its own ``spin moment" of magnitude $ \pm \hbar / 2 $

\qquad\qquad\qquad and intrinsic magnetic moment of magnitude
$ {| e | \hbar} / {2 \cm c} $}.
\end {list}

\section {Spin and Pauli equation}
After the emergence of the Schr\"odinger's theory in 1926, a new contradiction arose
with the Stern--Gerlach experiment since
silver atoms were prepared in a {\it spherically symmetric state}, which was known from spectroscopy. But the last term in the Schr\"odinger operator (\ref {HAB2}) cancels such states, since $ B \cdot \hat L $ is a generator of rotations around the vector $ B $.
Therefore,
there should be no first-order splitting!

In the language of quantum numbers, it looks like this:
in the $ s $-state, the angular momentum $ l = 0 $ under the action of a magnetic field is conserved, and
the multiplicity  of such a state is $ 2l + 1 = 1 $!

In 1927, Pauli proposed to modify the Schr\"odinger equation to double  the multiplicity of all stationary states in accordance with the two-digit quantum number 
introduced by Pauli  in his theory of periodic system  (Section  \ref {BPM}).
For this purpose, he introduced the two-component wave
functions
$ \psi (x, t) \! = \! (\psi_1 (x, t), \psi_2 (x, t)) \! \in \! \Co ^ 2 $ and added
to the energy operator (\ref {HAB})
a new {\it spin term} with coefficients corresponding to the Goldsmith--Uhlenbeck conjecture.
 

Namely,  formula (\ref {MH}) means that
the magnetic moment $ M $ appends the value $ -B \cdot M $ to  electron energy in the
magnetic field.
Therefore,
the last term in (\ref {HAB}) means that an electron  
has the magnetic moment $ M = \langle \psi, \fr e {2 \cm c} \hat L_3 \psi \rangle $,
in the $ \psi $ state 
due to its angular momentum $ L = \langle \psi, \hat L_3 \psi \rangle $. 
Therefore,
\be \la {MLeS}
 M = \fr {e} {2 \cm c} L,
\ee
which coincides with (\ref {MLe2}). 
Hence,
the Goldsmith--Uhlenbeck conjecture 
dictates for the Pauli two-component
wave functions
the replacement of $ M = \langle \psi, \fr e {2 \cm c} \hat L_3 \psi \rangle $
by $ M = \langle \psi, \fr e {2 \cm c} [\hat L_3 + 2 \hat s_3] \psi \rangle $,
where $ \hat s_3 = \fr \hbar 2 \si_3 $, and $ \si_3 $ is one of the Pauli matrices
\begin {equation} \la {tmm}
\sigma_1 = \left (\ba {rr} 0 & 1 \\1 & 0 \ea \right), \qquad \sigma_2 = \left (
\ba {rr} 0 & -i \\i & 0 \ea \right), \qquad \sigma_3 = \left (\ba {rr}
1 & 0 \\0 & -1 \ea \right).
\end {equation}
As a result, (\ref {HAB}) becomes the
 {\it Pauli equation}
\be \la {HABP}
i \hbar \dot \psi (x, t) = H_P \psi (x, t), \qquad
H_P = - \fr {\hbar ^ 2} {2 \cm} - \fr {e ^ 2Z} {| x |} - \fr e {2 \cm c} B_3 [\hat L_3 + 2 \hat s_3].
\ee
The eigenfunctions of the Pauli 
operator $ H_P $ are obviously expressed in terms of the eigenfunctions
$ H_B $,
\be
\psi_ {nlms} = \psi_ {nlm} \otimes e_s, \qquad s = \pm 1, \qquad e _ {+ 1} = (1,0), \quad e _ {- 1}. = (0,1) .
\ee
Namely, the 
spin term $ - \fr e {\cm c} B_3 \hat s_3 $ shifts the eigenvalues by $ \pm \fr {\hbar e} {2 \cm c} B_3 $
 in accordance with the Goldsmith--Uhlenbeck conjecture,
and for $ B = 0 $ the multiplicity of all stationary states
of the operator $ H_P$
doubles as compared to the operator (\ref {HAB}).

Thus  (\ref {nlmE2}) implies that the
eigenvalues of the Pauli operator $ H_P $ are numbered with the quantum numbers (\ref {nlms}), and $ s $ coincides with the eigenvalue of the 
operator $ 2 \hat s_3 $:
\be \la {nlmEB2}
\left \{
\ba {l}
 H ^ p \psi_ {nlms} = E_ {nms} \psi_ {nlms}, \, \,
 E_ {nms} = - \ds \fr {\cm e ^ 4} {2 \hbar ^ 2n ^ 2} - \fr {eB_3} {2 \cm c} \hbar [m + s]
 \\
 \\
 \hat L ^ 2 \psi_ {nlms} = \hbar ^ 2 l (l + 1) \psi_ {nlms}
 \\
 \\
\hat L_3 \psi_ {nlms} = \hbar m \psi_ {nlms}, \qquad
\hat s_3 \psi_ {nlms} = \hbar \fr s2 \psi_ {nlms}
\ea \right |
\quad
\ba {l}
 n = 1,2, \dots; \, \, \, l = 0,1, \dots, n-1;
 \\m = -l, \dots, l; \, \, \, s = \pm 1.
 \ea
 \ee
 Obviously, $ [\hat L ^ 2, H ^ p] = [\hat L_3, H ^ p] = [\hat s_3, H ^ p] = 0 $, so the observables
 $ \hat L ^ 2 (\psi ) $, $ \hat L_3 (\psi) $ and $ \hat s_3 (\psi) $ are conserved along the solutions of the Pauli equation
(\ref {HABP}).

  \br
{\rm
The sets of  possible energies
in (\ref {nlmEB2}) and (\ref {ELr42}) coincide
{\it with a fixed principal quantum number} $ n $
     because $ m + s = -n, \dots, n $.
}
\er

The Pauli equation (\ref {HABP}) allowed us to explain the experiments of Einstein--de Haas and
Stern--Gerlach, as well as many details of the
anomalous Zeeman effect, see \ci [Section 10.3] {K2016}.

  \subsection {Angular momentum and representations of rotation  group}
Commutation relations (\ref {LnH}) are equivalent to the commutation
of the dynamical group of the Schr\"odinger equation
$ W (t): \psi (0) \mapsto \psi (t) $ with
rotations $ R_n (\al): = e ^ {\al \pa_ \vp} $ of space $ \R ^ 3 $ around the $ x_n $-axis.
These rotations form a representation
of    the group $ SO (2) = U (1): = \{e ^ {i \al}: \al \in [0,2 \pi] \} $.
In particular, if the potentials are ``spherically symmetric,"
(\ref {Rn}) holds with respect to any axis of space. Then $ W (t) $ commutes with
representation
\be \la {Tt}
T (\theta) f (x): = f (R ^ {- 1} (\theta) x), \qquad f \in L ^ 2 (\R ^ 3), \qquad \theta \in \R ^ 3
\ee
of the whole group $ SO (3) = \{R (\theta): = e ^ {\theta r}: \theta \in \R ^ 3, | \theta | \le 2 \pi \} $, where
$ r = (r_1, r_2, r_3) $, and $ r_k $ are the generators of the rotations around the $ x_k $ axis:
\beqn \la {r1}
\left.
\ba {c}
r_1 = \left. \pa _ {\theta_1} \left (
\ba {rrr} 1 & 0 ~~ & 0 ~~ \\
0 & \cos \theta_1 & - \sin \theta_1
\\
0 & \sin \theta_1 & \cos \theta_1
\ea
\right) \right | _ {\theta_1 = 0} =
\left (\ba {rrr} 0 & 0 & 0 \\
0 & 0 & -1 \\
0 & 1 & 0
\ea
\right)
\\
\\
r_2 = \left. \pa _ {\theta_2} \left (
\ba {rrr} \cos \theta_2 & 0 & \sin \theta_2 \\
0 & 1 & 0
\\
- \sin \theta_2 & 0 & \cos \theta_2
\ea
\right) \right | _ {\theta_2 = 0} =
\left (\ba {rrr} 0 & 0 & 1 \\
0 & 0 & 0 \\
-1 & 0 & 0
\ea
\right)
\\
\\
r_3 = \left. \pa _ {\theta_3}
\left (
\ba {rrr} \cos \theta_3 & - \sin \theta_3 & 0 \\
\sin \theta_3 & \cos \theta_3 & 0
\\
0 & 0 & 1
\ea
\right) \right | _ {\theta_3 = 0} =
\left (\ba {rrr} 0 & -1 & 0 \\
1 & 0 & 0 \\
0 & 0 & 0
\ea
\right)
 \ea \right |.
 \eeqn
In this case, the conserved angular momentum (\ref {pL}) equals to the quadratic form of the operator
  \be \la {dp3}
\hat L = i \hbar T '(0).
\ee
  Here $ T '(0): = \na_ \theta T (\theta) | _ {\theta = 0} $ is the  generator of the 
representation $ T (\theta) $, and so
    \be \la {dp32}
    T (\theta) = \exp (- \fr {i \theta} \hbar \hat L).
\ee
It is easy to see that, vice versa, for each representation  $ T (\theta) $ of the rotation group commuting with the dynamical group, the quadratic form of its generator $ T '(0) $  is conserved in time.
\bd \la {ram}
{\rm
For any representation  $ T (\theta) $ of the rotation group
$ SO (3) $
in the phase space $ L ^ 2 (\R ^ 3) $ of the Schr\"odinger equation
the quadratic form $ \langle \psi, T '(0) \psi \rangle $ is called by definition the {\it angular momentum}
of the state  $ \psi $, which corresponds to this representation.
}
\ed

\subsection {Covariance of the Pauli equation with respect to rotations}
In covariant with respect to rotations
form, the Pauli equation reads
\be \la {PaA0}
i \hbar \dot \psi (x, t) = H_P (t) \psi (x, t) =
\fr1 {2 \cm} [\si \cdot (-i \hbar \na \! - \! \ds \frac
ec \bA^\ext (x, t)) ]^ 2 \psi (x, t) + e A_0 ^ {\rm ext} (x, t) \psi (x, t),
\ee
where $ \si: = (\si_1, \si_2, \si_3) $.
Equivalently,
\be \la {PaA}
i \hbar \dot \psi (x, t) = \fr 1 {2 \cm} [-i \hbar \na \! - \! \ds \frac
ec \bA^\ext (x, t)] ^ 2 \psi (x, t) \! - \! \frac
{e} {\cm c} \hat s \! \cdot \!
 \bB^\ext (x, t) \psi (x, t) + e A_0 ^ {\rm ext} (x, t) \psi (x, t),
\ee
where
 $ \hat s = \fr \hbar2 \si $  is the {\it spin operator}, and $ B ^ {\rm ext} (x, t) = \rot \bA^\ext (x, t) $.
In particular, for the case of a uniform magnetic field $ (0,0, B_3) $, 
the last equation coincides with  (\ref {HABP}) when
 the quadratic terms with $ e ^ 2 / c ^ 2 $ are neglected.

Equation  (\ref {PaA0}) can be written in the Hamiltonian form as 
\be \la {ph}
i \dot \psi (x, t) = \fr12 D \cH_P (\psi), \qquad \cH_P (\psi): = \langle \psi, H_P \psi \rangle.
\ee
The covariance means that the Pauli equation
retains its shape after 
rotations $ x \mapsto y = Rx $ with $ R \in SO (3) $,
followed by
suitable  transform for
 the potentials, the wave function and for
the basis in the space   $ \Co ^ 2 $ of values of the wave function.
Namely, let us  denote for $ \theta \in \R ^ 3 $
 \be \la {RRR}
 \! \! \! \! \! \! \! \!
 \left.
 \! \! \! \ba {c}
R (\theta) = \exp (\theta r), \, \, \, \ds U (\theta) = \exp (-i \fr {\theta \si} 2), \, \, \, \si (\theta) = U (\theta) \si U (- \theta), \, \, \,
\hat s (\theta) = U (\theta) \hat s U (- \theta)
\\
\\
\bA^\ext_ \theta (y, t) = R (\theta) \bA^\ext (R (- \theta) y, t), \, \, \, \bA^\ext _ {\theta, 0} (y, t) = A_0 ^ {\rm ext} (R (- \theta) y, t)
\\
\\
B ^ {\rm ext} _ \theta (y, t) = R (\theta) B ^ {\rm ext} (R (- \theta) y, t),
\ea \right |,
\ee
, where
$ r = (r_1, r_2, r_3) $, and $ r_k $ denote generators (\ref{r1}).
\br
{\rm
The correspondance $R(\theta)\mapsto U(\theta)$ is a two-valued 
{\it projective spinor representation} of the rotation group $SO(3)$.
}
\er
For any solution of the equation (\ref {PaA0}), we define
  \be \la {psit}
 \psi_ \theta (y, t) = U (\theta) \psi (x, t) = U (\theta) \psi (R (- \theta) y, t).
 \ee
Let us show that
 $ \psi_ \theta $ is a solution of the same equation in which $ \bA^\ext (x, t), \bA^\ext_ {0} (x, t), \si $, $ \hat s $, $ B ^ {\rm ext} (x, t) $
are replaced by $ \bA^\ext_ \theta (y, t), \bA^\ext _ {\theta, 0} (y, t), \si (\theta), \hat s (\theta), \bB^\ext_ \theta (y, t) $, respectively.

For a proof, we express the right-hand side (\ref {PaA}) in terms of $ \psi_ \theta (y, t) $
and apply $ U (\theta) $ to both sides. Since $ x = R (- \theta) y $, we have
\be \la {nax}
\na_ {y ^ l}\psi_ \theta (y) = \fr {\pa x_k} {\pa y ^ l}\na_ {x_k} U (\theta) \psi (x) = U (\theta) R_ {kl} (- \theta) \na_ {x_k} \psi (x) =
U (\theta) R_ {lk} (\theta) \na_ {x_k} \psi (x)
\ee
since the matrix $ R $ is orthogonal. Equivalently,
$ \na_y \psi_ \theta (y) = U (\theta) R (\theta) \na_x \psi (x) $.
In other words,
\be \la {nak}
\na_x \psi (x) = U (- \theta) R (- \theta) \na_y \psi_ \theta (y), \qquad \psi (x) = U (- \theta) \psi_ \theta (y ).
\ee
Hence,
\beqn \la {naxy}
U (\theta) [\si \cdot (-i \hbar \na_x \! - \! \ds \frac ec \bA^\ext (x, t))] ^ 2 \psi (x) & = & [ R (- \theta)
(\si (\theta) \cdot (-i \hbar \na_y \! - \! \ds \frac ec R (\theta)
\bA^\ext (R (- \theta) y, t)))] ^ 2 \psi_ \theta (y)
\nonumber \\
\nonumber \\
& = & [\si (\theta) \cdot (-i \hbar \na_y \! - \! \ds \frac ec \bA^\ext_ \theta (y, t))] ^ 2 \psi_ \theta ( y)
\eeqn
since the rotation of $ R (- \theta) $ preserves the ``length of the vector"
$ \si (\theta) \cdot (-i \hbar \na_y \! - \! \ds \frac ec \bA^\ext_ \theta (y, t)) $.

It remains to note that the matrices $ \si_k (\theta) $ obviously satisfy the same
multiplication rules as $ \si_k $, and
therefore, they have the same form (\ref {tmm}) in some new basis of the space $ \Co ^ 2 $.
This follows similarly to the Pauli theorem for the Dirac equation \ci [Theorem 11.4] {K2013}.

\subsection {Conservation laws for the Pauli equation}
Energy and momentum for the Pauli equation (\ref {PaA})
are defined  by the same formulas (\ref{HSH}) and (\ref{pL})
as for  the Schr\"odinger equation, and the proof of their conservation
remains unchanged.

However, the definition and proof of conservation 
for  the angular momentum 
require significant modification. Namely,
commutation (\ref {LnH}) and the conservation law (\ref {dp2}) are now generally incorrect.

\bd Angular momentum for  the Pauli equation is defined as
   \be \la {dLc}
J_n (t): = \langle \psi (t), \hat J_n \psi (t) \rangle, \qquad \hat J_n: = \hat L_n + \hat s_n.
 \ee

  \ed
  
  \bl\la{amcP}
  Let  the 
  external potentials satisfy the
  invariance conditions (\ref {Rn}). Then  for any solution to the Pauli equation  (\ref {PaA}) 
   the angular mometum  
   (\ref {dLc}) is concerved,
   \be \la {Lc}
J_n (t)= \const, \qquad t\in\R.
 \ee
\el
  \begin{proof}
This follows, as above, from commutation
   \be \la {JnH}
[\hat J_n, H (t)] = 0.
 \ee
For proof, it suffices to verify that
  \be \la {J32}
  [\hat J_n, \hat s \cdot \bB^\ext (x, t)] = 0.
  \ee
Let us  consider $ n = 3 $ for example.
Then
  \be \la {com}
  \left \{
  \! \! \! \! \! \! \! \! \! \!
\!\!\!
 \ba {rcl}
    [\hat L_3, \hat s \cdot \bB^\ext (x, t)] & = & \sum_k \hat s_k [\hat L_3, B_k ^ {\rm ext} (x, t)] = \sum_k \hat s_k \hat L_3 B_k ^ {\rm ext} (x, t)
    \\
    \\
    \qquad [\hat s_3, \hat s \cdot \bB^\ext (x, t)] & = & i \hbar \hat s_2 B_1 ^ {\rm ext} (x, t) -i \hbar \hat s_1 B_2 ^ {\rm ext} (x, t)
    \ea \right |,
  \ee
because $ [\si_1, \si_2] = 2i \si_3 $ (+ cyclic permutations).
Here $ \hat L_3 B_3 ^ {\rm ext} (x, t) \equiv 0 $, since the vector field $ B ^ {\rm ext} (x, t) $ is invariant under rotations $R_3(\vp)$
around the $ x_3 $ axis, as well as
$ \bA^\ext (x, t) $. Therefore, substituting $ \hat L_3 = i \hbar \pa_ \vp $, we get
  \be \la {r33}
 [\hat J_3, \hat s \cdot \bB^\ext (x, t)] = i \hbar \hat s_1 [\pa_ \vp B_1 ^ {\rm ext} (x, t) -B_2 ^ {\rm ext} (x, t)] + i \hbar \hat s_2 [\pa_ \vp B_2 ^ {\rm ext} (x, t) + B_1 ^ {\rm ext} (x, t)].
 \ee
It remains to verify that coefficients at $ \hat s_1 $ and $ \hat s_2 $ on the right-hand side 
vanish. First we note that the rotational invariance of 
the vector field $ (B_1 ^ {\rm ext} (x, t), B_2 ^ {\rm ext} (x, t)) $, means 
that   radial and angular components
of this vector field do not depend on the angle $\vp$:
    \be \la {Jaug}
    \left \{
\! \! \! 
\ba {rcl}
   B_r ^ {\rm ext} (x, t) = \cos \vp B_1 ^ {\rm ext} (x, t)) + \sin \vp B_2 ^ {\rm ext} (x, t) =f (r, x_3, t)
   \\
   \\
   B_ \vp ^ {\rm ext} (x, t)) = \sin \vp B_1 ^ {\rm ext} (x, t)) - \cos \vp B_2 ^ {\rm ext} (x, t) =g (r, x_3, t)
   \ea \right |
    \ee
    The differentiation in $ \vp $ gives
    \be \la {pavp}
     \left \{
\! \! \! 
\ba {rcl}
   \cos \vp B_1 ^ {\rm ext} (x, t)) + \sin \vp B_2 ^ {\rm ext} (x, t)
    - \sin \vp \, \pa_ \vp B_1 ^ {\rm ext} (x, t)) + \cos \vp \, \pa_ \vp B_2 ^ {\rm ext} (x, t) \equiv 0
   \\
   \\
  \sin \vp B_1 ^ {\rm ext} (x, t)) - \cos \vp B_2 ^ {\rm ext} (x, t)
+ \cos \vp \, \pa_ \vp B_1 ^ {\rm ext} (x, t)) + \sin \vp \, \pa_ \vp B_2 ^ {\rm ext} (x, t) \equiv 0
   \ea \right |.
    \ee
This easily implies the annihilation of the coefficients in  (\ref {r33}).
\end{proof}
  
 \bc
 {\rm
i) For the Pauli equation with {\it uniform magnetic field} $ B = (0,0, B_3) $
the operator (\ref {HABP}) commutes with the operator of  the ``orbital
angular momentum" $ \hat L_3 $ and with the
operator of the ``spin momentum" $ \hat s_3 $. Therefore, the quadratic forms of these operators
are conserved in time.
  \smallskip \\
ii) 
This commutation and conservation generally  does not hold 
for  magnetic fields  invariant with respect to rotations around the $ x_3 $ axis.
On the the hand, 
the quadratic form of the sum $ \hat J_3: = \hat L_3 + \hat s_3 $ is conserved by (\ref {JnH}). For {\it spherically
symmetric potentials}, the quadratic forms of the operators $ \hat L ^ 2 $ and $ \hat J ^ 2 $ are 
 also
conserved,
and their eigenvalues 
$ J_3, L ^ 2, J ^ 2 $
  are equal to
 \be \la {jj}
 \left \{
\!\!
 \ba {l}
 L ^ 2 = \hbar l (l + 1), \quad J ^ 2 = \hbar j (j + 1), \quad J_3 = \hbar m
 \\
 \\
\quad l = 0,1, \dots;
 \quad j = 0, \fr12,1, \fr32, .....; \quad m = -j, \dots, j
 \ea \right |,
 \ee
which is consistent with (\ref {nlmEB2}) in the case of a uniform magnetic field. 
The expressions for
$ J ^ 2 $ and $ J_3 $
stem from the general theorem on 
decomposition of tensor products  of
representations, applied to $ R (\theta) \otimes U (\theta) $, see  \ci {vdW1974}.
 }
 \ec
  \brs
  {\rm
i) 
Let  external potentials be {\it spherically symmetric}. Then
commutation (\ref {JnH}) implies the commutation of dynamical group $ W (t) $ with 
    \be \la {TtU}
    T_p (\theta): = \ds e ^ {- \fr i \hbar \theta \hat J} = \ds e ^ {- \fr i \hbar \theta \hat L} \ds e ^ {- \fr i \hbar \theta \hat s} = T (\theta) U (\theta),\qquad \theta\in\R^3,
    \ee
where $U (\theta)$ is 
defined in (\ref{RRR}).
In this case
     \be \la {psit2}
 \psi_ \theta (y, t): = U (\theta) \psi (R ^ {- 1} (\theta) y, t) = T_p (\theta) \psi (\cdot, t)
 \ee
according to (\ref {dp32}),
and the conserved angular momentum (\ref {Lc}) corresponds (up to a factor) to this representation in the sense of  Definition
\ref {ram}.
\smallskip \\
ii) For small $ \theta $, $ T_p (\theta) $  is the 
single-valued representation of the rotations $R(\theta)$, but
its analytic continuation for large $ \theta $ is a two-valued representation, as well as
 the spinor representation $ U (\theta) $.
\smallskip \\
iii)
The conservation of the momentum and angular momentum 
in quantum dynamics do not play an important role in quantum theory. 
On the other hand, the corresponding  commutation relations 
(\ref {LnH}) and (\ref {JnH})
play  a crucial role in the calculation of  eigenvalues of related  quantum observables, --
for example, in the calculation of  the anomalous Zeeman effect in \ci [section 10.3] {K2013}.

}
\ers

\section{Coupled nonlinear Maxwell--Schr\"odinger equations}\la{s16}

The Maxwell equations in the entire space of $ \R ^ 3 $ in
the Heaviside--Lorentz units read  (\ci [p. 781] {Jackson}) 
\be \la {mhl}
\dv \bE (x, t) = \rho (x, t), \, \, \, \rot \bE (x, t) = - \fr 1c \dot \bB (x, t), \, \, \dv \bB (x, t) = 0, \, \, \rot \bB (x, t) = \fr 1c [\bj (x, t) + \dot \bE (x, t)].
\ee
The second and third equations imply the Maxwell representations
$ \bB (x, t) = \rot \5 \5 \bA (x, t)$ and 
$ \bE (x, t) = - \fr 1c \dot \bA (x, t) -\na A ^ 0 (x, t). $
Then
in the Coulomb gauge $ \dv \bA (x, t) \equiv 0 $ 
the Maxwell equations (\ref {mhl}) are equivalent to the system
\be \la {Ma}
\fr 1 {c ^ 2} \ddot \bA (x, t) = \De \bA (x, t) +\fr1c P\bj (x, t), 
\quad \De A^0(x,t)= -\rho (x, t),\qquad x \in \R ^ 3,
 \ee
where 
 $ \rho (x, t) $ and $ \bj (x.t) $ are the charge and current  densities, respectively,
and $P$ denotes the orthogonal projection onto free-divergent vector fields  
from the Hilbert space $L^2(\R^3)\otimes\R^3$.
\smallskip

The Schr\"odinger equation (\ref{S}) describes the evolution of a wave function in a
given external Maxwell field
with potentials $\bA^\ext(x,t)$ and $A_0^\ext(x,t)$. On the other hand, the charge and current densities
(\ref{rho}), (\ref{j}) generate their ``own Maxwell field" with potentials satisfying the 
Maxwell equations 
(\ref{mhl-1}). Hence, for a self-consistent description,
these potentials  $\bA(x,t)$ and $A^0(x,t)$
should be added to the external potentials in the Schr\"odinger equation (\ref{S}). Thus,  
the coupled Maxwell--Schr\"odinger equations
  read
  as (cf. \ci{NW2007})
\be \la {MS}
\!\!\!\!\!\!
\left\{\ba{rcl}
\fr 1 {c ^ 2} \ddot \bA (x, t)\!\!\! &\!\!=\!\!&\!\!\! \De \bA (x, t) +\fr1c P\bj (x, t),
\quad \De A^0(x,t)= -\rho (x, t)
 \\
\\
 i \hbar \dot \psi (x, t)\!\!\!&\!\!=\!\!&\!\!\!
\fr 1 {2m} [- i \hbar \na\!-\! \ds \frac ec (\bA ( x, t)\!+\!\bA_\ext  ( x, t))]^2 \psi (x, t) \!+\! e (A ^ 0  (x, t)\!+\!A ^ 0_\ext (x, t)) \psi (x, t)
\ea\right|\, x \in \R ^ 3,
\ee
where $\bA_\ext ( x, t)$ and $A^0_\ext ( x, t)$ are some external Maxwell potentials,
$e<0$ is the electron charge and $c$ is the speed of light in a vacuum.
The coupling is defined by expressing the charge and current densities
in the wave function:
\be \la {rj}
\rho (x, t) = e | \psi (x, t) | ^ 2, \qquad \bj (x, t) = \fr e \cm \rRe \{\ov \psi (x, t) [- i \hbar \na- \fr ec (\bA_\ext  ( x, t)+\bA ( x, t))] \psi (x, t) \}.
\ee
These densities  satisfy the continuity identity
\be\la{cce}
\dot \rho (x, t) + \dv \bj(x, t) \equiv 0.
\ee
The system (\ref {MS})
is formally Hamiltonian, with the
Hamilton functional (which is the energy up to a factor)
\be\la{enc}
\cH(\bm{\Pi},\bA,\psi,t)=
\fr12[\Vert \fr1c \bm{\Pi}\Vert^2
+\Vert \rot\bA\Vert^2]+(\psi, H_\bA(t)\psi),
\ee
where 
$\Vert\cdot\Vert$ stands for the norm  in the real Hilbert space
$L^2(\R^3)\otimes\R^3$ and 
the brackets
$(\cdot,\cdot)$ stand for the inner product in 
$L^2(\R^3)\otimes\Co$.
The  Schr\"odinger magnetic operator
$$
H_\bA(t):=
\fr 1 {2m} [- i \hbar \na- \ds \frac ec (\bA ( x, t)+\bA_\ext  ( x))]^2
+e(\fr12 A^0(x,t)+A^0_\ext (x,t)),
$$
where $A^0(x,t):=(-\De)^{-1}\rho(\cdot,t)$ and $\rho(x,t):=e|\psi(x,t)|^2$.
The system (\ref{MS})  can be written in the Hamilton form as
\be\la{MSH}
\left\{\ba{l}
\fr 1{c^2} \dot\bA(t)=D_{\bm{\Pi}}\cH(\bm{\Pi(t)},\bA(t),\psi(t),t), 
\,\qquad
\fr 1{c^2} \dot{\bm{\Pi}}(t)=-D_\bA\cH(\bm{\Pi}(t),\bA(t),\psi(t),t)
\medskip\\
 i\hbar\dot\psi(t)=\fr12 D_\psi\cH(\bm{\Pi}(t),\bA(t),\psi(t),t),
\ea\right|,
\ee
taking into account that $(\psi eA^0,\psi)=(A^0,\rho)=
( (-\De)^{-1}\rho,\rho)$, and hence, 
$ D_\psi( \psi eA^0,\psi)=4eA^0\psi$.
Therefore, the energy is conserved in the case
of static external  potentials
\be\la{stat}
 \bA_\ext(x,t)\equiv\bA_\ext(x),\qquad
 A^0_\ext(x,t)\equiv A^0_\ext(x).
 \ee
 For instance, in the case of an atom,  $A^0_\ext(x)$ is the nucleus Coulomb potential, while
 $\bA_\ext(x)$ is  the vector  potential of the   nucleus magnetic field.
 On the other hand, the total charge $Q(t):=\ds\int\rho(x,t)dx$ formally 
is conserved for arbitrary  time-dependent  external potentials.

   \smallskip
      
   The Hamiltonian (\ref {enc}) is invariant with respect to the action of the group $ U (1) $,
\be \la {U1-2}
(\bA (x), \bm {\Pi} (x), \psi (x)) \mapsto (\bA (x), \bm {\Pi} (x), \psi (x) e ^ { i \theta}), \qquad \theta \in (0,2 \pi).
\ee
This invariance implies the
charge continuity equation (\ref {cce}) by
the general Noether theorem on invariants \ci [Section 13.4.3] {K2013}.
One can also check (\ref {cce}) by direct differentiation \ci [Section 3.4] {K2013}.
Moreover, for any solution $ \bA (x, t), \bm {\Pi} (x, t), \psi (x, t) $
the functions $ \bA (x, t), \bm {\Pi} (x, t) $, $ \psi (x, t) e ^ {i \theta} $ are also a solution.

\br
{\rm
The
existence of global solutions to the
Cauchy problems for
systems (\ref {MS}) in the entire space $ \R ^ 3 $ without external potentials was proved in \ci {GNS1995}
for all finite energy initial states  (\ref {enc}).
The uniqueness of the solutions has so far been proved only in narrower classes of functions,
   \ci {NW2007, BT2009}.
}
\er

Taking into account the electron spin, the Maxwell-Schr\"odinger system (\ref {MS}) should be replaced
by the  Maxwell--Pauli system with current density
\be \la {jrp}
j (x, t) = \fr em \rRe \si \{\ov \psi (x, t) \si \cdot [-i \hbar \na- \fr ec (\bA ( x, t) + \bA_\ext  (x, t))] \psi (x, t) \}.
\ee
where $ \si: = (\si_1, \si_2, \si_3) $ are the Pauli matrices.
In this case, the Schr\"odinger equation is replaced by the Pauli equation
$$
i \hbar \dot \psi (x, t) = H_P (t) \psi (x, t) =
\fr1 {2 \cm} [\si \cdot (-i \hbar \na \! - \! \ds \frac
ec (\bA (x, t)) + \bA_\ext  (x, t))] ^ 2 \psi (x, t) + e (A^0 (x, t) + A^0_\ext  (x, t)) \psi (x, t),
$$
and
the Schr\"odinger operator $ H_ \bA (t) $ in the 
Hamiltonian (\ref {enc})  must be replaced by the 
Pauli magnetic operator 
$$ H_ \bA ^ P (t): = \fr1 {2 \cm} [\si \cdot (-i \hbar \na \! - \! \ds \frac
ec (\bA (x, t)) + \bA_\ext  (x, t)] ^ 2 + e (\fr12A^0 (x, t)+A^0_\ext (x, t)). 
$$

   \br The system (\ref{MS})
was introduced essentially by Schr\"odinger   in his first articles \ci{Schr1926},
and it
underlies the entire theory of laser radiation
\ci{SZ1997}.
\er

\section {Quantum postulates   and  Maxwell--Schr\"odinger equations}

The Maxwell--Schr\"odinger equations (\ref{MS})
suggest a novel interpretation for basic quantum postulates: 
\smallskip
 

I. Transitions between quantum  stationary orbits.
 \smallskip
 

II. Wave-particle duality.
 \smallskip

III. Probabilistic interpretation.
 \smallskip\\
A rigorous dynamical description  of these postulates  is still
unknown.
This lack of theoretical clarity hinders the progress in the theory (e.g., in superconductivity and  in nuclear reactions),
and in
numerical simulation of many engineering processes (e.g., of laser radiation and quantum amplifiers) since a computer can solve dynamical equations but cannot  take into account  postulates.
\smallskip

Transitions between stationary orbits of atoms and
accompanying  radiation  
(Postulate I) 
were postulated by Bohr in 1913. 
The wave-particle duality (Postulate II) was conjectured by de Broglie in 1923,
and the probabilistic interpretation (Postulate III) was introduced by Born in 1927.
On the other hand, after the discovery of Schr\"odinger's quantum mechanics,
the question arose   on the validity of these postulates I--III in
new theory\,--\,this still remains an open problem.
These and other questions
have been frequently addressed in the 1920s and 1930s in discussions of Bohr, Schr\"odinger, Einstein and others \ci {B1949}.
However, a satisfactory solutions were not achieved.
We propose a novel approach to these problems relying on the
recent progress in the theory of attractors for Hamilton nonlinear PDEs.

The main goals of remaining part of these lectures 
are as follows:
\smallskip\\
i) To suggest a  relation of  these postulates to the theory of attractors.
\smallskip\\
 ii) To survey the related results
on global attractors for nonlinear Hamiltonian PDEs.
\smallskip\\
iii) To formulate   novel general
conjectures on the attractors.
\smallskip\\
iv)  To apply  these conjectures  to
 a mathematical interpretation
of    Postulates I--III in the context of coupled nonlinear
Maxwell--Schr\"odinger equations.

\section{Bohr's  postulates and attractors}

As a result, we expect the long-time asymptotics 
\be \la{ate1}
(\bA (x, t), \psi (x, t)) \sim (\bA_ \pm (x), 
e ^ {- i \om_ \pm t} \psi_ \pm (x)), \qquad t \to \pm \infty.
\ee
which holds 
in the $L^2$-norms and $H^1$-norms on every bounded region  of $\R^3$
{\it for all finite-energy solutions} to the  Maxwell--Schr\"odinger equations
(\ref{MS}).
\br
{\rm
Experiments show that the time of transitions (\ref {B1}) is of order $10^{-8}s$, though the asymptotics  (\ref {ate1}) requires an infinite time.
We suppose that this discrepancy can be explained by the fact that
 $10^{-8}s$ is the time when the atom emits an overwhelming part of the radiated energy.
}
\er

Such
asymptotics  are 
still open problems for the Maxwell--Schr\"odinger system
(\ref{MS}). On the other hand, similar asymptotics
are now proved for a number of model Hamilton nonlinear PDEs with symmetry groups $U(1)$. In next section we state a general conjecture which 
reduces to the asymptotics (\ref{ate1}) in the case of the Maxwell--Schr\"odinger system.

\bd
{\rm
{\it Stationary orbits} of the Maxwell-Schr\"odinger nonlinear system
(\ref {MS}) are
finite energy
solutions of  the form $ (A (x), e ^ {- i \om t} \psi (x)) $.
}
\ed
Existence of stationary orbits for the system (\ref {MS})
in the whole space was proved in \ci {CG2004} under conditions
\be \la {AAA}
\bA^{\rm ext} (x, t) \equiv 0, \qquad A^{\rm ext}_0 (x, t) = - \fr {eZ} {| x |}, \qquad \int | \psi_ \pm (x) | ^ 2 dx \le Z.
\ee

The asymptotics (\ref{ate1}) 
 mean {\it global attraction} to
the set of stationary orbits. We suggest similar attraction for the
Maxwell--Dirac, Maxwell--Yang--Mills and other
coupled equations.
In other words, we suggest to interpret quantum stationary
states as  points and trajectories lying  on the {\it global attractor} of the corresponding quantum dynamical equations.

\subsection{The Einstein--Ehrenfest paradox}
An instant
orientation of  the atomic magnetic moment
during $ \sim 10 ^ {- 4} s $ when turning on the magnetic field
in the Stern--Gerlach experiments 
caused the discussion in the ``Old Quantum Mechanics,"
because the classical model gave relaxation time
$ \sim 10 ^ 9 s $ taking into account the moment of inertia of the atom \ci {EE1922}.
In the  linear  Schr\"odinger's theory, this phenomenon also did not find a satisfactory explanation.
\smallskip

However, this instantaneous orientation is exactly in  line
 with  asymptotics (\ref {ate1}) for solutions to the
coupled Maxwell--Schr\"odinger  system.
Namely,
in the absence of a magnetic field,
the ground states (with a fixed charge) form a two-dimensional 
manifold.
When the magnetic field
 is turned on, the structure of the attractor 
(i.e., the set of corresponding stationary orbits $ (\bA, \psi) $)
instantly changes: 
the two-dimensional manifold bifurcates  
in two one-dimensional manifolds
 with a certain spin value. This bifurcation is not related to any 
 moment of inertia and  corresponds to the ``alternative
A" in the terminology of Einstein--Ehrenfest \ci {EE1922}:
{\it ``... atoms can never fall into the state in which they are quantized {\bf not fully}".}
 

\subsection{Attractors of dissipative and Hamiltonian PDEs }

Such interpretation of the Bohr transitions as a global attraction is rather natural.
On the other hand, the existing theory of attractors of {\it dissipative systems}
 \ci{L1944}--\ci{T1997} does not help  in this case, since
all funda\-men\-tal equations of quantum theory are {\it Hamiltonian}.
The global attraction for dissipative systems is caused by
energy dissipation. However, such a~dissipation
in the Hamilton systems is absent.

This is why we have developed in the 1990--2019s 
together with our collaborators
a novel theory
of global attractors for    {\it Hamilton}  PDEs, especially for application
to the problems of Quantum Theory.
Our results \ci{K1995a}--\ci{C2013}   for
the Hamilton equations rely on  energy radiation, 
 which
irrevocably carries the energy  to infinity and plays the role of  energy dissipation.
 A brief
survey of these results can be found in Section \ref{sattr}, and a detailed survey in \ci{K2016}.

The results  obtained so far indicate
an explicit {\it correspondence} between the type of 
 long-time asymptotics of finite energy solutions and  the
symmetry group  of the equation.
We formalize this correspondence in our general conjecture (\ref{at10}).

\section{Conjecture on attractors of $G$-invariant Hamilton nonlinear PDEs}
Let us consider
 $G$-invariant Hamilton nonlinear PDEs of type
 \be\la{dyn}
\dot\Psi(x,t)=F(\Psi(x,t)),\qquad x\in\R^3,\quad t\in\R,
\ee
with a Lie  symmetry group $ G $.
Here $\Psi(\cdot, t)$ belongs to the Hilbert space $L^2(\R^3)\otimes\R^d$,
and  the Hamilton structure means that
$F(\Psi)=J D \cH(\Psi)$, where $J^*=-J$.
The
$G$-invariance means that $ F (g \Psi) = gF (\Psi) $ for all  states
$ \Psi $ and all transformations $ g \in G $  (more precisely, $ g $ belongs to a representation
of the Lie  group $G$).
In this case, for any solution $ \Psi (t) $ to
equations (\ref {dyn}) the trajectory  $ g \Psi (t) $ is also a solution.

Let us note that
the theory of elementary particles deals systematically with the
symmetry groups
$ SU (2) $, $ SU (3) $, $ SU (5) $, $ SO (10) $ and other, and
$ G:=SU (4) \times SU (2) \times SU (2) $ is
the symmetry group of
``Grand Unification", see \ci {HM1984}.
\smallskip\\
{\bf Conjecture  A.} {\it
For ``generic'' $G$-invariant equations (\ref{dyn}),
any finite energy solution $ \Psi (t) $  admits a long-time asymptotics
\begin {equation} \label {at10}
\Psi (t) \sim e ^ {\hat g_ \pm t} \Psi_ \pm, \qquad t \to \pm \infty,
\end {equation}
where the generators $ \hat g_ \pm $
belong to the corresponding Lie algebra
$ \mathfrak {g} $ {\rm (}more precisely, $ \hat g_ \pm $ belong to a representation
of $ \mathfrak {g} ${\rm )}, and the above asymptotics holds
in some local seminorms.}
\smallskip

 In other words, all
$G$-orbits form a global attractor for  ``generic'' $G$-invariant Hamilton nonlinear
 PDEs of type (\ref{dyn}).
 This conjecture
 is a generalization of  rigorous results  \ci {K1995a}--\ci {K2016} obtained
 since 1990
 for  a list of model equations of type (\ref{dyn}) with three basic symmetry groups:
 the trivial group, the group of  translations, and the unitary  group $U (1) $.
 We give a brief survey of these  results in Section \ref{sattr}.
\smallskip\\
For the case of Maxwell--Schr\"odinger system (\ref{MS})
with the symmetry group $U(1)$,
the conjecture (\ref{at10}) reduces to the asymptotics (\ref{ate1}).
\smallskip\\
{\bf Empirical evidence.}
Conjecture (\ref{at10}) agrees with the Gell-Mann--Ne'eman theory of baryons  \cite{GM1962, Ne1962}.
 Namely, in 1961
Gell-Mann and Ne'eman  suggested  the symmetry group $SU(3)$ and other ones
for the strong interaction of baryons  relying on the discovered parallelism between empirical data
for the baryons,
and the ``Dynkin scheme''
of the Lie algebra $\mathfrak{g}=su(3)$ with $8$ generators (the famous ``eightfold way").
This theory resulted in
the scheme of quarks and in the
development of the
quantum chromodynamics \cite{HM1984},
and in the prediction of a new baryon
with prescribed values of
its mass and decay products. This particle  (the $\Omega^-$-hyperon)
was promptly discovered experimentally \cite{omega-1964}.
The elementary particles seem to describe long-time asymptotics
of quantum fields. Hence,
the empirical correspondence between elementary
particles and generators of the Lie algebras
presumably gives an
evidence in favour of our general conjecture
(\ref{at10}) for equations with  Lie symmetry groups.
\smallskip

Conjecture (\ref{at10}) suggests
to define stationary ``$G$-orbits'' for equations (\ref{dyn}) as solutions of the type
\begin {equation} \label {at10a}
\Psi (t) = e ^ {\hat g t} \Psi, \qquad t \in \R,
\end {equation}
 where $\hat g$
belongs to the corresponding Lie algebra
$ \mathfrak {g} $ (more precisely, $ \hat g $ belong to a representation
of the Lie algebra $ \mathfrak {g} $).
 This definition  leads to the corresponding
``$ \mathfrak {g} $-eigenvalue problem"
\begin {equation} \label {at10b}
\hat g \Psi = F (\Psi).
\end {equation}
In particular, for  the linear Schr\"odinger equation
with the symmetry  group  $U (1) $,
stationary orbits are solutions of the form $ e ^ {i \om t} \psi (x) $,
where $ \om \in \R $ is an eigenvalue of the Schr\"odinger operator, and $ \psi (x) $ is the corresponding eigenfunction. 
However,
Conjecture  (\ref{at10}) fails for linear equations, i.e., linear  equations are exceptional,
 not ``generic''!
In the case of the symmetry group $ G = SU (3) $, the generator
(``eigenvalue") $ \hat g $ is
$ 3 \times 3 $ -matrix, and solutions (\ref {at10a}) are quasiperiodic in time.
  

\section {Results on global attractors for nonlinear Hamilton PDEs}
\la{sattr}

Here we describe
rigorous results \ci {K1995a}--\ci {C2013}
obtained since 1990
on the corresponding asymptotics
for a number of Hamiltonian nonlinear partial differential equations of type (\ref{dyn}).
We give  only a brief listing of the results,
see the details in \ci {K2016}. The results obtained
confirm the existence of finite-dimensional attractors
in the Hilbert phase space, and demonstrate
an explicit correspondence between
the long-time asymptotics and  the symmetry group $ G $ of equations.
\smallskip

The results were obtained  so far for
model equations with
three basic groups of symmetry:
the trivial symmetry group
$ G = \{e \} $, the translation group $ G = \R ^ n $ for translation-invariant equations, and the unitary group $ G = U (1) $ for phase-invariant equations.
 
 
\subsection{Global attraction to stationary states}

\begin{figure}[htbp]
		\begin{center}
			\includegraphics[width=0.7\columnwidth]{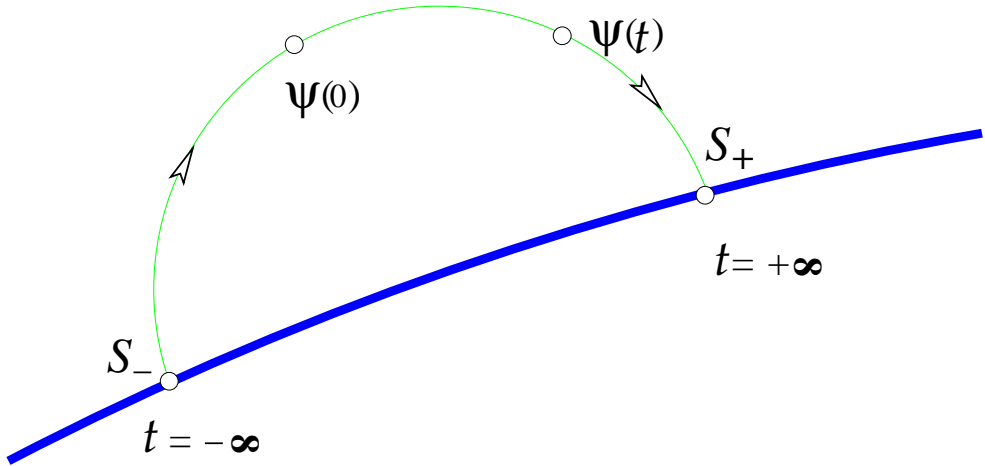}
			\caption{Convergence to stationary states.}
			\label{fig-1}
		\end{center}
	\end{figure}

{\bf For ``generic'' equations with  trivial symmetry group}
 the long-time asymptotics of all finite energy solutions
 is the convergence to  stationary states
 (see Fig.~\ref {fig-1})
 \begin {equation} \label {ate}
\psi (x, t) \to S_ \pm (x), \qquad t \to \pm \infty
\end {equation}
which was proved for a number of model equations
in \ci {K1995a}--\ci {KK2019b}: i) for a string coupled to nonlinear oscillators, ii) for a~three-dimensional wave equation coupled
to a  charged particle and for the Maxwell--Lorentz equations, and also
iii) for  the
wave equation, and the Dirac and Klein--Gordon equations
with concentrated nonlinearities.

Here $S_\pm(x) $ are some stationary
states depending on the considered trajectory $\psi(x,t) $,
and
the convergence holds in local seminorms of type $L^2(|x|<R)$ for
any $R>0$.
The convergence (\ref{ate}) in global norms (i.e., corresponding to
$R=\infty$) {\bf cannot hold} due to
energy conservation.

\bex {\bf Nonlinear Huygens Principle.}
{\rm
Consider solutions to  3D wave equation  with a unit propagation velocity and initial data with support in a ball $ | x | <R $. The corresponding
solution is concentrated in the spherical layers $ | t | -R <| x | <| t | + R $. Therefore, the solution converges everywhere to zero as
$ t \to \pm \infty $, although its energy remains constant.
This convergence to zero is  known as the
{\it  strong  Huygens principle}. Thus,  the attraction to stationary states
(\ref {ate}) is a generalization of this Huygens principle to nonlinear equations.
The difference is that for a linear wave equation the limit behind the wave front 
 is always zero, while for nonlinear equations
the limit can be any stationary solution.
}
\eex
The proofs
in \ci {KSK1997} and \ci {KS2000}
rely on the relaxation of the acceleration
\be \la {rel}
\ddot q (t) \to 0, \qquad t \to \pm \infty.
\ee
Such relaxation has been known for a long time
in classical electrodynamics as the “radiation damping”, but it was first proved in \ci {KSK1997} and \ci {KS2000} for charged relativistic particle in a scalar field
and in the Maxwell field under the {\it Wiener Condition} on the particle charge density.
This condition is an
analogue of the ``Fermi Golden Rule'', first introduced by Sigal in the context of nonlinear wave- and Schr\"odinger equations \ci {Sigal}.
The proof of the relaxation (\ref{rel}) relies
on a novel application of the Wiener Tauberian theorem.

\subsection{Global attraction to solitons}

{\bf For ``generic'' translation-invariant equations},
the long-time asymptotics of all finite energy solutions
is the convergence to solitons
\begin {equation} \label {att}
\psi (x, t) \sim \psi_ \pm (x-v_ \pm t), \qquad t \to \pm \infty,
\end {equation}
where the
convergence holds in local seminorms
{\it in the comoving frame of reference}, that is, in
$ L ^ 2 (| x-v_ \pm t | <R) $ for any $ R> 0 $.
Such soliton asymptotics were  proved in
\ci {KS1998}--\ci {IKS2004b}
for three-dimensional wave equation coupled
to a charged particle and for the Maxwell--Lorentz equations.
These results gave the first rigorous proof of  the ``radiation damping'' in classical electrodynamics, which has been an open problem for about 100 years.
The proofs
in \ci {KS1998} and \ci {IKM2004}
rely on  variational properties of solitons and their orbital stability, as well as on the  relaxation of the acceleration (\ref {rel})
under the Wiener
condition on the particle charge density.

More accurate soliton asymptotics in global norms
with several solitons were first discovered
in 1965 by Zabuzhsky and Kruskal in
numerical simulation of the Korteweg--de Vries equation (KdV):
it is the decay to solitons
\be \la {attN}
\psi (x, t) \sim \sum_k \psi_ \pm (x-v ^ k_ \pm t) + w_ \pm (x, t), \qquad t \to \pm \infty,
\ee
where
$ w_ \pm $ are some dispersion waves.
 In \ci {KMV2004} the  results of
numerical simulation were presented to confirm the soliton asymptotics (\ref {attN})
with many solitons
for 1D {\it relativist-invariant} nonlinear wave equations.
Later on, such asymptotics
were proved by the method of {\it inverse scattering problem}
for nonlinear
\textbf {integrable} Hamiltonian
translation-invariant equations (KdV, etc.)
in the works of Ablowitz, Segur, Eckhaus, van Harten and others \cite {EvH}.
A trivial example is provided by  the d'Alembert equation
$ \ddot \psi (x, t) = \psi '' (x, t) $,
for which any solution reads
$ \psi (x, t) = f (x-t) + g (x + t) $.

\subsection{Global attraction to stationary orbits}

\begin{figure}[htbp]
	\begin{center}
		\includegraphics[width=0.9\columnwidth]{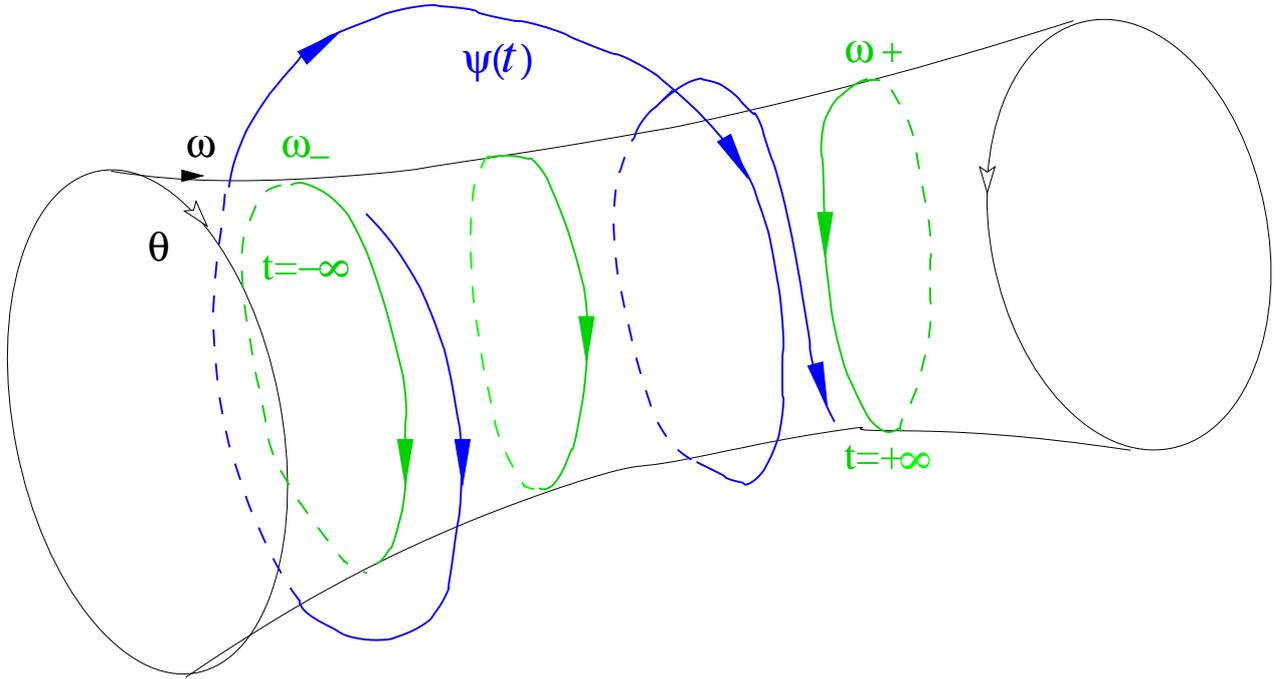}
		\caption{Convergence to stationary orbits.}
		\label{fig-3}
	\end{center}
\end{figure}

{\bf For ``generic'' equations with unitary symmetry group $G=U(1)$}, the long-time asymptotics are global attraction
to ``stationary orbits'' (see Fig.~\ref{fig-3})
\begin{equation}\label{atU}
	\psi(x,t)\sim\psi_\pm(x) e^{-i \omega_\pm t} , \qquad t \to \pm \infty;
\end{equation}
they were proved in
\ci{KK2006}--\ci{C2012} for the Klein--Gordon and Dirac equations coupled to
$U(1)$-invariant nonlinear oscillators, and in
\ci{C2013},  for discrete in space and time difference approximations
of such coupled systems, i.e., for the corresponding difference schemes.
\smallskip

The global attraction was proved under the assumption that the equations
are ``strictly nonlinear''. For linear equations, the attraction can fail
if the discrete spectrum consists at least of two points.

\begin{remark} \label {rg}
{\rm Let us comment on the term {\it generic} in the
results of the previous section and in Conjecture (\ref {at10}). Namely, this conjecture
means that the asymptotics (\ref {at10}) hold for all solutions for
an {\it open dense set} of $G$-invariant equations.
\smallskip \\
i) For example, asymptotics
 (\ref {ate}), (\ref {att}), (\ref {atU})
hold under appropriate conditions, which define some ``open dense set'' of
$G$-invariant equations with three types of the symmetry group $G$: either
under the  Wiener condition or under the strict nonlinearity condition, etc.
The asymptotics
may break down if these conditions fail\,---\,this corresponds to some ``exceptional''
equations: for example, asymptotics   (\ref {atU})  break down for the linear Schr\"odinger equations with at least two different eigenvalues.
\smallskip \\
ii) The general situation is the following. Let a Lie group $ G_1 $ be a (proper) subgroup of some  larger Lie group $ G_2 $. So,
the $ G_2 $-invariant equations
form an ``exceptional subset'' among all $ G_1 $-invariant equations, and
the corresponding asymptotics (\ref {at10}) may be completely different.
For example, the trivial group $ \{e \} $ is a subgroup in $ U (1) $ and in $ \R ^ n $, and asymptotics
(\ref {att}) and (\ref {atU}) may differ significantly  from (\ref {ate}).

}
\end{remark}

\section{De Broglie' wave-particle duality}
In 1923, de Broglie 
suggested in his PhD
to
identify the beam of particles with a harmonic wave:
\begin{equation}\la{dB1}
\!\!\!\!
\mbox{\it a beam of  particles with moment  $p$ and energy
$E=\fr{p^2}{2\cm}$ }
\Lra\,\,
 \psi(x,t)=Ce^{i(kx-\om t)},\,\,\,(p,E)=\hbar (k, \om).
\end{equation}
This identification was suggested as a counterpart to the Einstein
corpuscular treatment of light as a 
 beam of photons.
The duality (\ref{dB1})
 was  the key source for the Schr\"odinger quantum mechanics.

We suggest a mathematical description
of  the wave-particle duality 
relying on a {\it generalization  of the conjecture} (\ref{at10}) for the case of
 translation-invariant Maxwell--Schr\"odinger system (\ref{MS}) {\it without external
potentials},
i.e., 
$
\bA^\ext(x,t)\equiv 0$, $A_0^\ext(x,t)\equiv 0.
$
 In this case, the Schr\"odinger equation of (\ref{MS}) becomes
 \be\la{S0} 
 i \hbar \dot \psi (x, t)=
\fr 1 {2m} [- i \hbar \na\!-\! \ds \frac ec \bA ( x, t)]^2 \psi (x, t) \!+\! e A ^ 0(x, t) \psi (x, t)
,\quad x \in \R ^ 3.
 \ee
Now  
  the symmetry group 
  of system (\ref{MS}) 
  becomes $G=\R^3\times U(1)$, and our general conjecture (\ref{at10}) should be strengthened similarly to
 (\ref{attN})
\be \label {SA}
\bA (x, t) \sim
\displaystyle \sum \limits_ {k} \bA_ \pm ^ k (x-v ^ k_ \pm t) + \bA_ \pm (x, t), \,\,\,
\psi (x, t) \sim
\displaystyle \sum \limits_ {k} \psi_ \pm ^ k (x-v ^ k_ \pm t)
e ^ {i \Phi_ \pm ^ k (x, t)} + \psi_ \pm (x, t), \quad
t \to \pm\infty
\ee
for  each finite energy solution, where $ \bA_ \pm (x, t)$ and $\psi_ \pm (x, t)$
 stand for the corresponding {\it dispersion waves}.
 The solitons (traveling wave solutions)
$( \bA (x-v t), \psi (x-v t))$
 for (\ref{MS}) were constructed in
 \ci{CG2004}.
 These asymptotics suggest to treat the solitons as electrons and
provisionally correspond to the {\it reduction (or collapse)} of wave packets.
\smallskip
  
 The asymptotics (\ref{SA})
suggest
a mathematical description
of  the wave-particle duality
{\it under several assumptions}.
Namely, 
let us consider the wave function $\psi(x,t)=Ce^{i(kx-\om t)}$
as initial data. 
Then initially  the 
corresponding charge and current densities 
\be\la{roj}
\rho(x,t)=e|\psi(x,t)|\equiv e|C|^2,\qquad
\b\bj(x,t)=\fr e\cm \rRe\{\ov\psi(x,t)[-i\hbar\na \psi(x,t) ]\}\equiv  \fr e\cm |C|^2\hbar k
\ee
 are uniform.
 Hence, initially   the Maxwell field $\bE(x,t)$ and $\bB(x,t)$ vanish, as well as 
 the potentials $\bA(x,t)$ and $A^0(x,t)$. 
 Therefore,
 the Schr\"odinger equation
  (\ref{S0}) implies that 
    the wave function $\psi(x,t)=Ce^{i(kx-\om t)}$   satisfies
    initially
the free Schr\"odinger equation which implies
$
\hbar\om=\fr {\hbar^2k^2}{2\cm}.
$
\smallskip

Further {\it we expect} that 
 the {\it   space-averaged}
charge, the momentum and the energy densities  do not depend on time due to
the corresponding conservation laws.
Then the density of the electrons (solitons)  should be   $n=|C|^2$.
Similarly,  the density of
momentum and energy of the solitons
should be, respectively,
$
\ds P=-i\hbar \ov\psi(x,t)\na\psi(x,t)=\hbar |C|^2k.
$
and
$
\ds\cE=\ds\fr{\hbar^2}{2\cm}|\na\psi(x,t)|^2
=\fr{\hbar^2}{2\cm}|C|^2k^2.
$
Finally, it is natural to assume that the velocities
of the solitons should be identical by translation
homogeneity.
Then the momentum $p=P/n$ and energy $E=\cE/n$  of one electron are equal, respectively, to $  p=\hbar k$ and
 $E=\fr{\hbar^2k^2}{2\cm}=\fr{p^2}{2\cm}=\hbar\om$
  which agrees with (\ref{dB1}).
\smallskip

\section{Born's
probabilistic interpretation}

In 1927, Born suggested the probabilistic interpretation of the wave function:

 {\it The probability of detecting an electron at a point $ x $ at the time $ t $ is proportional to $ | \psi (x, t) | ^ 2 $.}
\medskip

We suggest below an  interpretation of
this postulate  relying on asymptotics  (\ref{SA}) for the coupled Maxwell--Schr\"odinger equations (\ref{MS}).
However, the corresponding
rigorous justification  for the nonlinear equations (\ref{MS}) is still an open problem.

 \subsection {Diffraction of electron beams}\la{spi2}

 Born proposed the probabilistic interpretation
    to describe the diffraction experiments
of Davisson and Germer of the 1924--1927s.
In these experiments,
 the electron beam was scattered by a nickel crystal, and the reflected beam was fixed on a
photo-film. The resulting images are similar to
``Lauegrams'', 
which were
first obtained in 1912 by the method of Laue.
Later on, such experiments were also carried out with transmitted electron beams
scattered by a thin gold and platinum crystalline films
(G.\,P.~Thomson, Nobel Prize 1937).
Recently Bach \& al.
{\it for the first time
observed double-slit diffraction} of electrons \ci {BPLB2013},
which is the first realization for electronic waves
of the classical Young's experiment of 1803.
 

The
electron diffraction was considered for a long time  as a paradoxical phenomenon incompatible with the concept of
a point elementary particle.
On the other hand, the diffraction phenomena are inherent properties of
the linear Schr\"odinger equation
as was
commonly recognized for a long time \ci{AB1959}--\ci{Eskin2013}.
We show in the next section (see also \ci{Kdif2019}) that
there is  a fine {\it quantitative}
agreement of solutions to the linear Schr\"odinger equation
for two-slit screen
with the results of recent diffraction experiments
\ci{BPLB2013}.
Namely, according to the {\it principle of limiting amplitude}, the diffracted wave admits the asymptotics
\be \la {dif}
\psi_d (x, t) \sim a_ \infty (x) e ^ {- i \om t}, \qquad t \to \infty,
\ee
where $ \om $ is the frequency of the incident wave. The calculation
of the diffraction amplitudes $ a_ \infty (x) $ in the next section
using {\it the  Kirchhoff approximation}
demonstrates that the maxima  of
 $ a_ \infty (x) $ on
 the screen agree very well with those
of the diffraction pattern in
experiments \ci {BPLB2013}.
Thus, the diffraction of electron beams
finds  a~natural basis in the linear Schr\"odinger  theory.

\subsection{Discrete registration of electrons}

However, in 1948 the probabilistic interpretation received
new content and confirmation
after the experiments of Biberman, Sushkin and Fabrikant
\ci {BSF1949}.
In these experiments with very low electron beam intensity,
the diffraction pattern was created as an averaging of random
discrete registration of individual electrons.
Later on, similar experiments were carried out by
Chambers, Tonomura, Frabboni,
Bach \& al. \ci {Chambers60, Tonomura89, FGP07, BPLB2013}.
To explain this phenomenon, there are at least two possibilities,
which are both related to random fluctuations:
\smallskip\\
{\bf i)}  {\it Random interaction with counters.}
One possibility to explain the discrete registration is a
random triggering either a) of registration counters located at the screen points,
or
b) of atoms of the photo emulsion.
We suppose that the probability of triggering
is proportional to the  current, which is given by
\be \la {ja}
j (x,t) = \fr e \cm \rRe \{\ov \psi (x, t) [- i \hbar \na \psi (x, t)] \} \approx \fr {e \hbar} \cm k | a_ \infty (x) | ^ 2,\qquad t\to\infty.
\ee
by (\ref {dif}) and
according to (\ref {rj}) with $ \bA _ {\rm ext} (x,t)= 0 $.
Let us note that  we set $ \bA _ {\rm ext} (x,t)= 0 $
since there is no  external fields between the scatterer screen and the screen
of observation. The term with $ \bA (x,t) $ in (\ref {rj})  is also neglected since it is
relatively small.

\begin{figure}[htbp]
\begin{center}
\includegraphics[width=0.5\columnwidth]{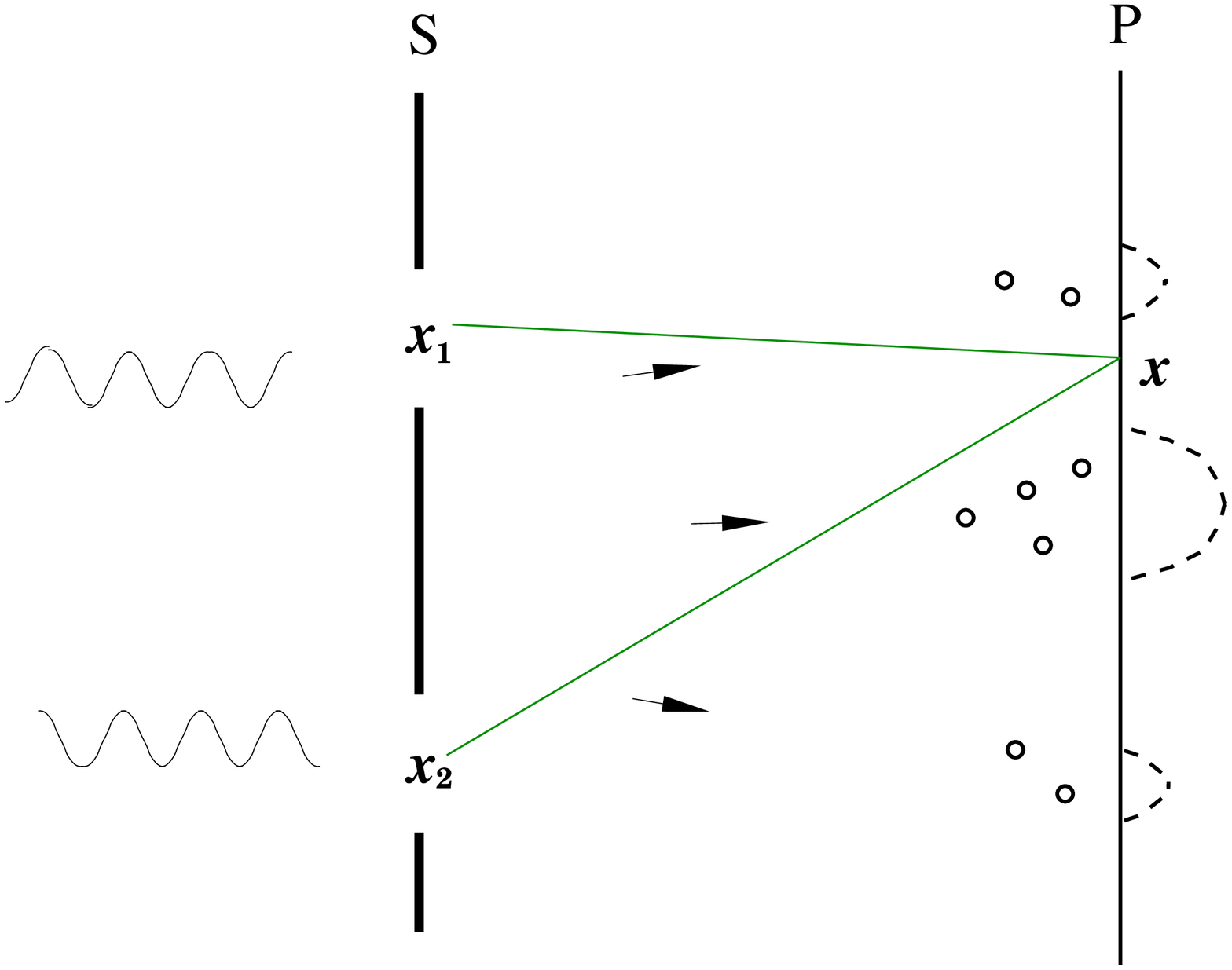}
\caption{Diffraction\index{diffraction} by double-slit.} \label{fT0}
\end{center}
\end{figure}

Therefore, the averaged diffraction pattern should have maxima
at the screen points with {\it maximal electric current},
which coincide with the screen points with {\it maximal amplitude}
$ |a_ \infty (x)| $ by (\ref {ja}).
 This coincidence is confirmed  in  \ci {Kdif2019}
by calculations  of $ a_ \infty (x) $  and by comparison  with
experiments \ci {BPLB2013}.
Thus, the discrete registration of electrons also admits
 an interpretation in the linear Schr\"odinger
theory.
  \smallskip \\
{\bf ii)} {\it Random reduction of wave packets.}
Another possibility to explain the discrete registration
is the {\it soliton-conjecture} (\ref{SA}) for translation-invariant
Maxwell--Schr\"odinger  system (\ref{MS}).
This conjecture is inspired by the asymptotics
(\ref{attN}), which was proved for
translation-invariant integrable nonlinear PDEs.
 Respectively,
we suppose that the decay  (\ref{SA})
should hold between the scatterer screen and the screen of observation, where the {\it external fields vanish},
and hence, the system (\ref{MS}) is translation-invariant,
see Fig.~\ref{fT0}.
Such a~decay into solitons  should be considered as a
random process, as it is subject to microscopic fluctuations.

An averaged  registration rate of electrons
at a point of the screen
should be proportional to the current (\ref{ja})
if the contribution of the dispersion waves $\psi_\pm$
is negligible.
This follows from the charge conservation law. Therefore, the averaged diffraction pattern again should have maxima
at the screen points with  {\it maximal amplitude}
$ |a_ \infty (x)| $.
Thus,
 this treatment of the discrete registration of electrons
 requires
 the {\it soliton-conjecture} (\ref{SA}) for nonlinear
Maxwell--Schr\"odinger  equations.
 \bigskip


\section{On diffraction  of electrons and  Aharonov--Bohm shift}\la{sdif}

 \begin{abstract}
We calculate  the amplitude of diffraction for the electron beams in the framework of   the Kirchhoff
approximation applying  the limiting amplitude and the limiting absorption principles, and  the Sommerfeld 
radiation condition.
The Aharonov--Bohm Ansatz and the corresponding shift of the diffraction pattern  
are justified
for small magnetic field by
a novel reduction to short-range magnetic potential and subsequent application of the 
Agmon--Jensen--Kato stationary scattering theory.

\end{abstract}

\subsection{Introduction}

The diffraction of the electron beams was observed first by Davisson and Germer in the 1924--1927s 
\ci{DG1927} (see also \ci[Section 5.2]{K2013}).
In these first experiments,  the electron beam was scattered by the nickel crystal and the reflected beam was fixed on a
film. The resulting images are similar to X-ray scattering patterns
(lauegrams), first obtained in 1912 by the method of Laue. 
Later, such experiments were also carried out with transmitted electron beams passing through thin crystal films of gold and platinum
 (Thomson 1927). Only  recently  Bach \& al carried out for the first time the  two-slit diffraction of electrons  \ci {BPLB2013}.

The influence of the ``excluded magnetic field" on the
  diffraction process,    
  which was
  first predicted in 1949 by 
Ehrenberg and Siday \ci{ES1949},  became widely discussed after 
  the work of
  Aharonov and Bohm \ci{AB1959}, and was confirmed experimentally in  1960 by Chambers
\ci{Chambers60}. 
There is a huge amount of works  concerned with  various aspects of the A-B effect:
the shift of the diffraction pattern, the change in the  scattering matrix and in the discrete spectrum etc.
The survey and the references can be found   in  \ci{Eskin2015, OP1985,PT1989}.
\smallskip

The diffraction of electrons  was initially recognized  as a paradoxical phenomenon incompatible with the concept of a
point elementary particle 
(now the  non-point nature of electrons and of other elementary particles is  well established experimentally). 
The paradoxical appearance of the ``particle diffraction"  inspired the introduction   of the  probabilistic interpretation of the wave function   by Born in
 1927 (though the actual need of the  probabilistic interpretation arises at a novel stage -- when interpreting a discrete
registration of individual electrons in the Biberman--Suchkin--Fabrikant  experiments of 1949 with super-weak intencities of  falling beams).

 On the other hand,  the electron diffraction completely agrees with the wave nature of the electron, which was formalised by 
the Schr\"odinger  theory  based on the wave-particle duality of de Broglie.
We apply the Fresnel-Kirchhoff theory to the calculation of the  diffraction amplitude and check that the formula agrees satisfactory
with    the results of  recent experiments  \ci {BPLB2013}. Our paper aroses exactly because of the possibility of such a comparison.

Similarly, the main topic of discussions on A-B shift is always the distinguished role 
of the 
Maxwell potentials in quantum theory, which is considered as the testimony of the nonlocal nature 
of the interaction of electrons with the Maxwell field.
However, these 
discussions suggest implicitly again
that electrons are point particles 
that do not pass through the region of the magnetic field.
\smallskip

In present paper, we calculate the diffraction amplitude 
of the transmitted electron beam for general aperture in the plane screen.
The incident electron beam is  described by 
a plane wave. 
We assume i) the limiting amplitude principle and ii) the Sommerfeld radiation condition for the limiting amplitude,
 though both these assumptions are not rigorously justified for the plane scatterers.
These assumptions allows us 
to
express the  {\it limiting amplitude} via its Cauchy data
by the  Kirchhoff approach which relies on the Green integral formula.
The Green function is selected via the ``limiting absorption principle"  providing
 the Sommerfeld  radiation condition, which 
cancels the integral over the large half-sphere.

The  Cauchy  boundary data of the limiting amplitude
 are approximated by the Cauchy data of the incident wave. This approximation 
 is commonly recognized in Optics after   Kirchhoff works on the diffraction. 
 It is  still not justified rigorously for the Schr\"odinger equation,  though it is broadly used 
 in the physical literature, see the survey  \ci[pp. 390--419]{OP1985}  and the references therein. 
  The resulting formulas for the diffraction amplitude include
a geometric factor known from the Fresnel--Kirchhoff diffraction theory. 
We check that this approximation 
  agrees satisfactory with recent experiments (Bach \& al \ci{BPLB2013}) in the particular case
  of two-slit aperture, which confirms to some extent the validity of the  Kirchhoff  approximation. 
\smallskip

We extend these calculations to the case of a localized ``excluded magnetic field" to justify
the Aharonov-Bohm effect.
Many  theories of this effect rely on a gauge transform,
which reduces the perturbed Schr\"odinger equation
to the unperturbed one. However, this fact alone does not imply 
automatically
the same correspondence 
(known as {\it Aharonov-Bohm Ansatz} \ci{BW2011})
between 
the diffraction amplitudes
of the free and perturbed equations. 
Such correspondence was justified in \ci{BW2009cmp, BW2009jmp, BW2011}
 in the context of  scattering of finite-energy wave packets moving in external electric and magnetic potentials. 
Different appearances of the A-B effect were studied in \ci{Eskin2013,Eskin2015,Helffer, Ruij}.
 
We justify the A-B Ansatz for the corresponding diffraction amplitudes {\it  in the case of small magnetic fields}
in the original framework 
of  scattering of plane waves \ci{AB1959} applying the Kirchhoff approximation.
Such a scattering problem in the same approximation
was  solved   in \ci{AB1959}  for the case when the 
 aperture is the whole plane and the magnetic field is supported  by an infinite straight line (a ``magnetic string"). 
 
 We assume  the limiting amplitude principle and  the Sommerfeld radiation condition for the limiting amplitude.
 Our main result is an expansion of the magnetic Green function 
{\it satisfying the Sommerfeld radiation condition}
in the case of 
 {\it  a small magnetic field}. 
  This expansion relies on a novel
 reduction to the case of a {\it continuous  potential with rapid  decay} and
subsequent application of the Agmon--Jensen--Kato stationary scattering theory.
 
 The Sommerfeld condition specifies the long-range asymptotics of the Green function 
and plays the role of  boundary conditions at infinity.  
The role of boundary conditions in the A-B effect
 was pointed out  by Peshkin \ci [p. 21] {PT1989}. 
 
\br
{\rm
Our calculation shows that  the A-B shift holds for small magnetic field in the {\it zero order approximation}, while
in {\it higher order approximations} the change of the diffraction pattern provisionally does not reduce to a shift.
}
\er



\subsection{The electron diffraction}

Schematically, the experiment on the diffraction of a transmitted electron beam is shown on Fig. \ref{fT0}. The incident  beam of electrons with energy $E$ falls
from the left on the scattering plane $ x_3 = 0 $
 with the apertures $Q$. The transmitted waves interfer on the observation screen  $P$,
located in the half-plane
$\Om:=\{x\in\R^3: x_3>0\}$  (see Fig. \ref{fT0}). 

\noindent{\bf The diffraction problem.} The incident beam of free electrons is described by the wave
\be\la{vpx}
\vp_{in}(x,t)=a_{in} e^{i(kx_3-\om t)},\qquad  \om=E/\hbar,  
\ee
in accordance with the de Broglie wave-particle duality. It
satisfies
the {\it free
relativistic Klein--Gordon equation}
\be\la{KG0}
\frac
1{c^2}[i\hbar\pa_t]^2\psi(x,t)=[(-i\hbar\na)^2+
\cm^2c^2]\psi(x,t).
\ee
This equation is derived from the relativistic energy formula
$\fr{E^2}{c^2}=p^2+\cm^2c^2$ by the 
canonical quantization
  $E\mapsto  \hat E:= i \hbar \pa_t$, $p\mapsto \hat p:=-i \hbar \na$.
For  $E\approx mc^2$,
this relativistic energy is close to the nonrelativistic one,
\be\la{reap}
 E=\sqrt{p^2c^2+m^2c^4}\approx mc^2(1+ \fr{p^2}{2m^2c^2})=mc^2+\fr{p^2}{2m}.
 \ee
The  canonical quantization leads to the free
  Schr\"odinger equation with the shift
  \be\la{reap}
 i\hbar\dot\vp(x,t)=-\fr{\hbar^2}{2m}\De\vp(x,t)+
 mc^2\vp(x,t).
 \ee
 This shift is easily removed by the gauge transformation
  $\vp(x,t)=\psi(x,t)e^{-i\om_0t}$,
 where $\om_0=mc^2$. Now $\psi(x,t)$ 
 satisfies the standard free
  Schr\"odinger equation
 \be\la{S02}
i\hbar\partial_t\psi(x,t)=-\frac {\hbar^2} {2\cm}
\De\psi(x,t),
\ee
 and the incident wave is replaced by
  $\psi_{in}(x,t)=a e^{i(kx_3-\om' t)}$, where $\om'=\om-\om_0$ and 
 \be\la{kom}
 \hbar\om'=\fr {\hbar^2 k^2}{2m}.
  \ee
Let us   choose a coordinate system in which the scattering screen
   $S$ lies in the plane
  $x_3=0$, 
  and the electron beam falls from the region
  $x_3<0$. 
  The external potentials vanish everywhere in the connected domain $\R^ 3\setminus S$, and hence
   the
    diffraction problem
   reduces to solving the free Schr\"odinger equation
   (\ref{S02}) 
  in $\R^3\setminus S$
  with the  ``boundary condition"
   \be\la{reap1}
  \psi(x,t)- \psi_{in}(x,t)\to 0,\qquad x_3\to-\infty.
    \ee
    We {\it assume} 
      the  {\it limiting amplitude principle}  to hold
        \be\la{reap12}
 \psi(x,t)\sim a_\infty (x)e^{-i\om' t},\qquad t\to\infty.
 \ee
 It is well established 
 for the  Schr\"odinger equation with a short-range potential 
 and for boundary value problems outside a {\it smooth bounded
 domain}
 \ci{Eidus69, Eidus89},
 \ci[Ch. 28]{KK2012}.
 The limiting amplitude  satisfies  the free Schr\"odinger equation (\ref{S02})
   \be\la{reap2}
H_0(k)a_\infty(x):= [\De+k^2]a_\infty(x)=0,\qquad   x\in\Om
 \ee
 due to (\ref{kom}).
 The solution can be expressed
  through its Cauchy data on $\pa\Om$ using the Green formula
 \be\la{Gr20}
a_\infty(x)=\int_{y_3=0+} [G_0(x,y)\pa_{y_3} a_\infty(y)-\pa_{y_3}G_0(x,y) a_\infty (y)]dy, \qquad x\in\Om,
 \ee
 where  the Green function $G_0(x.y)$ is the integral kernel 
 of the free Helmholtz operator $H_0(k+i0)$ on $\R^3$:
 \be\la{aGf}
H_0(k) G_0(\cdot,y)=\de(x-y),\qquad x,y\in\Om;\qquad G_0(x,y)=-\ds\fr {e^{ik|x-y|}}{4\pi |x-y|}.
 \ee
 The integral over the large halh-sphere 
 $\{y\in\Om: |y|=R\}$
 vanishes as $R\to\infty $ due to the Sommerfeld radiation condition
 \ci[formulas (34.5)]{Som53}, which holds 
 for this Green function  and for the amplitude
 \be\la{SRC}
 \left\{\ba{rclrcl}
a_\infty(y)&=&\cO(|y|^{-1}),   & \pa_{|y|}a_\infty(y)&=&ika_\infty(y)+o(|y|^{-1})
\\
\\
 G_0(x,y)&=&\cO(|y|^{-1}), & \pa_{|y|}G_0(x,y)&=&ikG_0(x,y)+o(|y|^{-1})
 \ea\right|,\quad |y|\to\infty.
 \ee
 The last conditions are  obvious
   for the Green function (\ref{aGf}).
   We {\it assume} these conditions also
 for the limiting amplitudes, since they  are
 well established for 
 the  Schr\"odinger equation with a short-range potential 
 and for boundary value problems outside a smooth bounded
 domain, see \ci{Eidus69,Eidus89} and also formulas (28.1),  (27.7) and (27.8) in  \ci{KK2012}.

 \smallskip
 
\noindent{\bf The Kirchhoff approximation.}
 The main idea of the  {\it Kirchhoff method} \ci{BW} is the approximation of the
  Cauchy data of solutions by the same of the incident wave (\ref{vpx})
   \be\la{reap2k}
 a_\infty(y)|_{y_3=0+}\approx D_{in}(y),\quad \pa_{y_3}a_\infty(y)_{y_3=0+}\approx N_{in}(y), 
  \ee
  where
    \be\la{reap3}
 D_{in}(y):= \left\{\ba{ll} a_{in},&x\in Q\\0&y\in S\ea\right|,
 \qquad    N_{in}(y):=\left\{\ba{ll} ika_{in},&y\in Q\\0&y\in S\ea\right| .
 \ee
  For the Green function (\ref{aGf}) we have for $x_3> 0$ and $y_3 = 0$
 \be\la{pan22}
G_0(x,y)=-\fr{1}{4\pi}\fr{e^{iks}}{s} ,\qquad
\pa_{y_3} G_0(x,y)=-\fr{ik}{4\pi}\fr{e^{iks}}{s} \cos(y-x,e_3)+\cO(s^{-2}),\qquad s:=|x-y|,
\ee
where $e_3:=(0,0,1)$.
Substituting  these approximations 
into the  Green formula (\ref{Gr20}),
we get
 \be\la{reap5}
a_\infty(x)=-\fr{ika_{in}}  {(4\pi)^2} \int_Q \fr{e^{ik|x-y|}} { |x-y|}(1+\cos\chi)dy,\qquad x\in\Om,
 \ee
where  $\chi$ is the ``Fresnel angle of diffraction" between the vectors $e_3$ and $ x-y$, see \ci[Fig. 8.1]{BW}.
This result can also be obtained from the Kirchhoff--Fresnel formula (17) of
\ci[Section 8.3]{BW}
for the case of an incident spherical wave with a point source  $y$,
when it is removed to infinity
 $y_3\to-\infty$.

 \brs
 {\rm
 i) 
 Physically
 the Kirchhoff approximation  (\ref{reap2k})   means that the electron density at the points of the aperture $Q$
 equals $|a_{in}|^2$, while the density of the momentum  equals  $\ov\psi_{in}\na\psi_{in}=k|a_{in}|^2$. These
 assumptions are commonly recognised in physical literature, see the survey  \ci[pp. 390--419]{OP1985}  and the references therein. In particular,
 this approximation is used  in the paper \ci{AB1959}.
 However, this approximation is not sill justified rigorously for the scattering problem with the ``boundary condition" (\ref{reap1}).
 \smallskip\\
 ii) Formulas of type (\ref{reap5})  can be found in almost all publications on the diffraction of  
 electrons, however only with  $\cos\chi\equiv 1$. 
 
 }
 \ers


\noindent{\bf Fraunhofer asymptotics.}
For bounded apertures $Q$,  the integral (\ref{reap5})
admits 
the  asymptotics
in each fixed direction
$x/|x|=(\xi_1,\xi_2,\xi_3)$  as $|x|\to\infty$,
\be\la{Fraun}
a_\infty(x)\sim -\fr{ika_{in}(1+\cos\ov \chi) e^{ik|x|}}  {(4\pi)^2|x|} \int_Qe^{-ik(\xi_1y_1+\xi_2 y_2)}dy,\qquad |x|\to\infty,
\ee
where $\cos\ov\chi=\xi_3>0$.  This asymptotics corresponds to the {\it 
Fraunhofer diffraction}
\ci[Section 8.3.3, formula (28)]{BW}.
In particular, for one circular aperture of radius $ R $
centered at $ y = 0 $ the
integral
(\ref {reap5}) has been calculated in 1835 by Airy 
\ci[Section 8.5.2, formula (13)]{BW},
\be\la{krFraun}
a_\infty(x)\sim -\fr{ika_{in}(1+\cos\ov \chi) e^{ik|x|}}  {(4\pi)^2|x|} \,\,\fr{2\pi r^2 J_1(kR\sin\ov\chi)}{kR\sin\ov\chi},\qquad |x|\to\infty,
\ee
where  $\sin\ov\chi= |(\xi_1,\xi_2)|$.

 \subsection{Comparison with the experiment}
The two-slit diffraction of electrons 
was observed experimentally first in 2013 (Bach \& al \ci {BPLB2013}): the
formula (\ref{reap5}) is in satisfactory agreement with the results of these
experiments. Namely,  
in the particular case, 
when the aperture $Q$ consists of two slits centered at the points $q_1, q_2\in S$,
 formula (\ref {reap5}) gives
 \be\la{dif}
 a_\infty(x)\sim \fr{e^{ik|x-q_1|}}{|x-q_1|}(1+\cos\chi_1)+\fr{e^{ik|x-q_2|}}{|x-q_2|}(1+\cos\chi_2),\qquad x\in\Om. 
 \ee
In these experiments
the distance between the planes $ S $ and $ P $ is $ D = 240 $ $ \mu $ m $ = 240.000 $ nm, the distance {\it between the centers of the slits} is $ 2d=|q_1-q_2| = 330 $ nm,
and the wavelength $ \lam = 50 $ nm. Then for $x\in P$
$$
|\cos\chi_j-1|\le |x-q_j|^2/D^2\le 1/240^2,\qquad |x-q_j|<1000 ~{\rm nm}. 
$$
Therefore, for such deviations, we can set $ \cos \chi_j \approx 1 $, and then the formula
(\ref {dif})  gives
\be\la{difcx}
 |a_\infty(x)|\sim 
 \fr 2D
 \Big| e^{ik[|x-q_1|-|x-q_2|]} +1\Big|. 
 \ee
 Let  points $q_1$, $q_2$ be symmetric with respect to the origin.
Choose
the $x_1$-axis   orthogonal to the plane of drowing, 
so that $ q_1 =(0,d,0)$ and  $q_2=(0,-d,0) $, where $d:=|q_1-q_2|/2$, see Fig. \ref{fABT}. Then 
 the amplitude (\ref{difcx}) has the maxima
at the points $x\in P$ with coordinates $x_2$ determined by the Bragg rule
\be\la{Bragg}\sqrt{(x_2-d)^2+D^2}-\sqrt{(x_2+d)^2+D^2}=n\lam,\quad n=0,\pm1,\dots
 \ee
 Expanding the square root in a Taylor series, we get the equation
  $2dx\approx  n\lam D$. For $n=0$ we get the central maximum $x_2=0$, and for
  $n=1$ we get the second maximum
 $x_2=\lam D/|q_1-q_2|=\fr{50}{330}240\approx 36,4$ $\mu$m. 
 An experimental value, as
  obtained in \cite{BPLB2013}, is
  approximately the
 $40$ $\mu$m
 \ci[Fig. S2 a]{BPLB2013data}.
 \br\la{ramp}
 {\rm
 Formula  (\ref{difcx}) satisfactory reproduces  positions of the maxima, but not their magnitudes. 
 On the other hand, 
 the picture \ci[Fig. S2 a]{BPLB2013data}  shows   complete coincidence of the measured diffraction amplitude with
  the graph obtained by numerical integration over paths. This integration in the limit should coincide with the exact solution of the 
diffraction problem 
(\ref{S02}), (\ref {reap1})
with the incident wave (\ref {vpx}).
 }
\er

\subsection {The Aharonov--Bohm shift}

The two-slit arrangement is superimposed by a~magnetic field
$B(x)$ concentrated in a closed  tube $\cT\subset\Om$ 
lying between the plane of scattering $S$ and the pane of observation $P$.
The tube
passes parallel to the  $x_1$-axis in the region near the aperture $Q$ and
encloses sufficiently far from the aperture. 
\begin{figure}[htbp]
\begin{center}
\includegraphics[width=0.5\columnwidth]{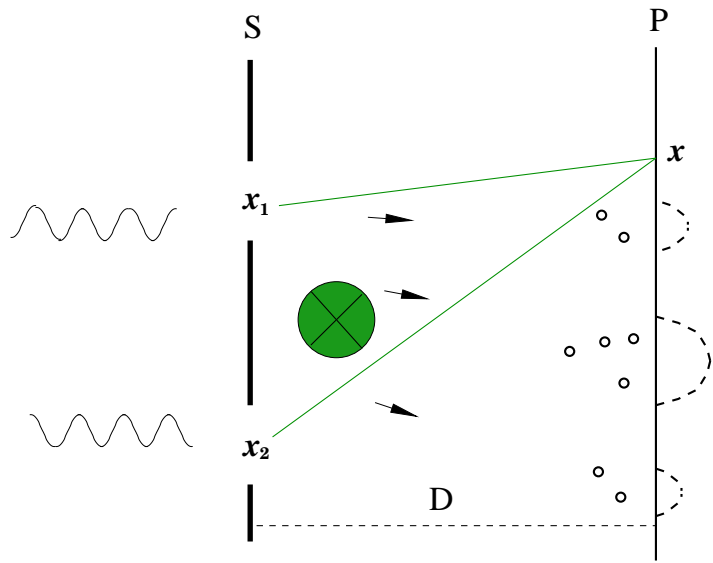}
\caption{Shift in magnetic field}
\label{fAB}
\end{center}
\end{figure}
In this case the diffraction pattern of the   set-up of Fig. \ref{fT0}
is shifted downwards.
This shift  was first 
experimentally observed by Chambers~\cite{Chambers60}.
The set-up was close to  that of Fig.~\ref{fAB}.
A complete  survey of experimental observations and various theoretical  interpretations can be found in \ci{PT1989}.
\smallskip

Let us calculate this AB-shift applying the 
Kirchchoff approximation  (\ref{reap2k}), (\ref{reap3})
of the Cauchy data.
Now the free Schr\"odinger equation (\ref{S02}) for $\psi(x,t)$
changes to the equation  
\be\la{S3}
i \hbar \dot \psi(x,t) =
\fr 1{2\cm}[-i\hbar\na- \ds\frac ec A(x)]^2\psi(x,t),
\ee
where $A(x)$ is the vector potential of the external
magnetic field  in the Coulomb gauge:
\be\la{Cg}
B(x)=\rot A(x),\qquad \dv A(x)=0,\qquad x\in\Om.
\ee
Respectively, 
the stationary Helmholtz equation  (\ref{reap2}) for the
limiting amplitude  changes to
\be\la{S4}
H(k)a_\infty^B(x):=
[\na+ \ds\frac e{i\hbar c}A(x)]^2a_\infty^B(x)
+k^2a_\infty^B(x)=0.
\ee
Now the Green formula   (\ref{Gr20}) becomes
\be\la{Gr21}
a_\infty^B(x)=\int_{y_3=0+} \Big[G(x,y)[\pa_{y_3} a(y)+ 2\ds\frac e{i\hbar c}A_3(y) a(y)]-\pa_{y_3}G(x,y) a (y)\Big]dy, \qquad x\in\Om,
 \ee
where the Green function  is now 
the integral kernel of the limit operator $H^{-1}(k+i0)$ on $\R^3$.
To justify this integral representation 
we will prove i) the existence (in a sense) of this limit  operator and  ii) the Sommerfeld radiation 
condition for $G(x,y)$ 
similar to (\ref{SRC}) for $G_0(x,y)$. We assume that this condition 
also holds for the limiting amplitude $a_\infty^B(x)$.
 The radiation conditions 
of type (\ref{SRC}) 
for 
$G(x,y)$ and  $a_\infty^B(x)$ 
cancel the 
 integral over the large half-sphere in  (\ref{Gr21})  as well as  in  (\ref{Gr20}).

 Further we will
substitute into  (\ref{Gr21}) the Kirchhoff approximation 
(\ref{reap2k}), (\ref{reap3})
of the Cauchy
data of the amplitude
and then we will check the A-B shift. We will accomplish  this program for a small magnetic field.

 

\subsection{Reduction to  short-range magnetic potential.}
\begin{figure}[htbp]
\begin{center}
\includegraphics[width=0.5\columnwidth]{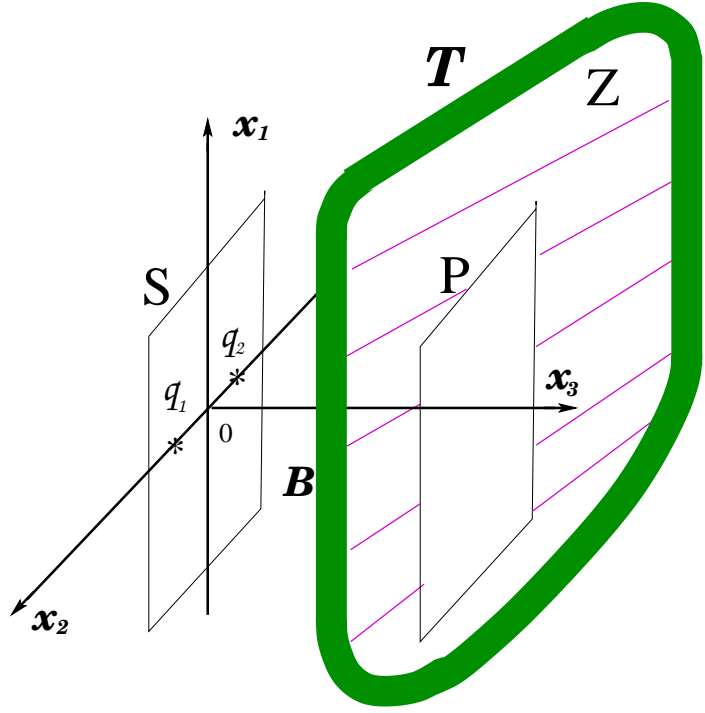}
\caption{Magnetic chain and the film.}
\label{fABT}
\end{center}
\end{figure}

 We plan to construct the magnetic Green function $G(x,y)$
 using
  the Agmon--Jensen--Kato 
 stationary scattering theory  \ci{Ag, JK}. 
 First we reduce the problem to magnetic potentials with compact support. This  is necessary  for
estimates (\ref{S12}) below.
 The potential $A(x)$ decays slowly. For example, 
 for an infinite cylindrical tube $\cT$   
\be\la{ABS}
\int_{\pa S}A(x)dx=\int_S B(x)dx=\Phi\ne 0
\ee
for any disk $S$ in $\Om$
if  its boundary embraces the tube  $\cT$. Hence, the
potential $A(x)$ decays slowly like $r^{-1}$, where $r$ is the distance from the tube. 
For a finite tube, the decay is faster but it is also insufficient for our purposes.
\br
{\rm 
There were many attempts to avoid the slow decay 
by constructing ``non-Stokesian potentials" $A$ supported by the tube $\cT$ \ci[pp 65--70]{PT1989}.
Then the ``Stokes identity" (\ref{ABS}) breaks down, which means that these potentials
 {\it do not satisfy the equation
 $B(x)=\rot A(x)$
 in the sense of distributions  in the region} $\Om$.
}
\er
To
reduce the problem to the case of potentials with rapid decay, 
 we choose the magnetic potential 
 solving the equations (\ref{Cg}) for all $x\in\R^3$ assuming $B\in H^1(\Om)$ and $\supp B\subset\cT$. Then we have 
 $\rot B=-\De A$, which implies that $A\in H^2_{\rm loc}(\R^3)\subset C(\R^3)$ by the Sobolev embedding theorem, and moreover,
 \be\la{Apot}
 \max_{|x|\le R} |A(x)|\le  C(R)\Vert B\Vert_{H^1(\R^3)},\qquad
 A(x)=\int_{\R^3}\fr{\rot B(y)dy}{4\pi|x-y|},\quad x\in\Om
 \ee
 for the potential $A(x)$ vanishing at infinity.
 Further, 
the magnetic field $B(x)=\rot A(x)$ vanishes outside the  tube $\cT$, and hence,
\be\la{S5}
\rot A(x)=0,\qquad x\in
\Om\setminus\cT.
\ee
Therefore, 
\be\la{S6}
A(x)=\na\phi(x),\qquad x\in
\Om\setminus(\cT\cup Z),\qquad \phi(x)=\ds\int_{\ga_\infty^B(x)} A(y)dy\in C^1(\R^3),
\ee
where $Z$ is a bounded surface in $\R^3$ (a ``film")
such that the domain $\ti\Om:=\Om\setminus (\cT\cup Z)$ is simply connected, see Fig. \ref{fABT},
while 
 ${\ga_\infty^B(x)} $ is any path in $\ti\Om$
connecting  $x$ with a fixed point.

Let us denote by $\ti\cT$ a bounded neighborhood of $\cT\cup Z$.
By (\ref{S6})
we can split the vector potential as 
\be\la{Abv}
A(x)=b(x)+\na \vp(x),\qquad x\in\Om,
\ee
 where 
\be\la{chivp}
\vp\in C^1(\Om),\qquad
\vp(x)=\phi(x)\,\,{\rm for}\,\, x\in \Om\setminus \ti\cT;\qquad b\in C(\Om),\quad\supp b\subset\ti\cT
\ee
Moreover,  the estimate (\ref{Apot}) implies that
\be\la{ABb}
\beta:=\max_{x\in\ti\cT} |b(x)|\le  C\max_{|x|\le R(\ti\cT)} |A(x)|\le  C_1\Vert B\Vert_{H^1(\R^3)}
\ee
where $R(\ti\cT)$ is the radius of a ball which contains the tube $\ti\cT$. 
The splitting (\ref{Abv}) allows us to factorize the operator $H(k)$ as follows:
\be\la{S7}
H(k)=e^{-\fr{e}{i\hbar c}\vp(x)} H_b(k)e^{\fr{e}{i\hbar c}\vp(x)},\qquad H_b(k):=[\na+ \ds\frac e{i\hbar c}b(x)]^2+k^2.
\ee
Now 
the  theory \ci{Ag, JK} implies the existence
of the limits
\be\la{S8}
H^{-1}(k+i0)=e^{-\fr{e}{i\hbar c}\vp(x)} H_b^{-1}(k+i0)e^{\fr{e}{i\hbar c}\vp(x)}
\ee
for all values of $k\in\R$, except a discrete set. 
The limit exists in the space of linear operators acting in suitable {\it weighted Sobolev spaces}, see Theorem 19.2 of \ci{KK2012}.
Then for the corresponding integral kernels
\be\la{S9}
G(x,y)=e^{-\fr{e}{i\hbar c}\vp(x)} G_b(x,y)e^{\fr{e}{i\hbar c}\vp(y)}.
\ee

\subsection{Asymptotic expansion for  a small magnetic field} 
Let us construct the series for  $G_b(x,y)$ which converges for a small magnetic field $B$.
First, $H_b(k):=H_0(k)+ b$,
where $b=\sum_{j=1}^3 b_j(x)\pa_{x_j}+b_0(x)$, and  $b_0(x)=\fr{e^2}{\hbar^2 c^2} b^2(x)$.
Second, formally,
\be\la{S110}
H_b^{-1}(k):=(1+ H_0^{-1}(k)b)^{-1}   H_0^{-1}(k)= H_0^{-1}(k)+R,
\ee
where 
 \be\la{S120}
R=\sum_{n=1}^\infty    R_n,\qquad R_n=
(-1)^n (H_0^{-1}(k)b)^n H_0(k)^{-1}.
\ee 
Respectively,
\be\la{S11}
G_b(x,y)=G_0(x,y)+\sum_1^\infty R_n(x,y),
\ee
where $R_n(x,y)$ is the integral kernel of the operator $R_n$.
For example,
  \be\la{R1}
 R_1(x,y)= \int_{\ti\cT} G_0(x,z)[\sum_{j=1}^3 b_j(z)\pa_{z_j}+b_0(z)]G_0(z,y)dz.
  \ee
Similar expressions hold for $R_n(x,y)$ with arbitrary $n\ge 1$. 
Hence,
each term $R_n(x,y)$ 
 is smooth for $x\ne y$ 
 similarly to  the Green function $G(x,y)$,
 and has the singularity $\sim |x-y|^{-1}$ near the diagonal $x=y$ only for $x,y\in\ti\cT$.

Let us note that in the representation of the limiting amplitude (\ref{Gr21}) we need the Green function only for $y\in Q$.
Moreover, the diffraction 
is observed for  sufficiently large $|x|$.
So
in all cases $x,y\not\in \ti\cT$. For such $x,y$ the integral representations of type (\ref{R1}) imply that
\be\la{S12}
| R_n(x,y)|\le C\fr{\Vert C_1\beta\Vert^{N+1}}{(|x|+1)(|y|+1)},\quad |\pa_{|y|} R_n(x,y)|\le C\fr{\Vert C_1\beta\Vert^{N+1}}{(|x|+1)(|y|+1)^2},\qquad x,y\not\in \ti\cT.
\ee
On the other hand, the derivation of the representation  (\ref{Gr21})  requires the 
 Sommerfeld radiation condition 
 \be\la{Som}
 G(x,y)=\cO(|y|^{-1}),\quad
 |y|(\pa_{|y|}-ik)G(x,y)\to 0,\qquad |y|\to\infty.
 \ee
The estimates (\ref{S12})е
imply that
  these conditions hold for each term $R_n(x,y)$ and hence
for  
    $G_b(x,y)$ 
  when $C_1\beta<1$.
Moreover, 
\be\la{Som2}
|y\pa_{|y|}\vp(y)|= |y A(y)|\to 0,\qquad |y|\to\infty,
\ee
which follows from  (\ref{Apot}) by the partial integration.
 Hence, the conditions  (\ref{Som}) hold by (\ref{S9}), 
 so 
 the integral representation (\ref{Gr21}) is justified for small    $ \Vert B\Vert_{H^1(\R^3)}$ by
 the inequality (\ref{ABb}).

\br\la{rB}
{\rm
We assume that $B\in H^1(\Om)$ to have $A,b\in C(\Om)$ which implies the
representations (\ref{S7}) and (\ref{S8}).
We suppose that similar representations  also hold for $B\in L^2(\R^3)$ --
however this is an open question.
}
\er

\subsection{The zero-order approximation}
In the zero order approximation,
the magnetic Green function reads
\be\la{S90}
G(x,y)\approx e^{-\fr{e}{i\hbar c}\vp(x)} G_0(x,y)e^{\fr{e}{i\hbar c}\vp(y)}.
\ee
Respectively,  the Cauchy data (\ref{pan22}) 
 change to
\be\la{Cd0}
\!\!\!\!\!\!\!
\left\{
\!\!
\ba{rcl}
G(x,y)&\!\!\!\!\sim&\!\!\!\! -e^{-\fr{e}{i\hbar c}  [\vp(x)-\vp(y)]}\fr{1}{4\pi}\fr{e^{iks}}{s} 
\\
\\
\pa_{y_3} G(x,y)&\!\!\!\!\sim&\!\!\!\!
-e^{-\fr{e}{i\hbar c}  [\vp(x)-\vp(y)]} \Big[ik \cos(y-x,e_3)+  \fr{e}{i\hbar c}     \pa_{y_3}  \vp(y)\Big] \fr{1}{4\pi}\fr{e^{iks}}{s}+\cO(s^{-2})
\ea\right|\,\,\,  y_3=0+.
\ee
Finally, 
for small magnetic fields 
the terms with $\fr e{i\hbar c}$ in (\ref{Gr21}) and (\ref{Cd0}) can be neglected with respect to $|k|$ 
(let us note that $|k|\sim 10^9 $ cm$^{-1}$ for $\lam=50$ pm as in the experiment of \ci{BPLB2013}).
Now the integral (\ref{Gr21})  becomes similar to (\ref{reap5}),
 \be\la{reap52}
a_\infty^B(x)\sim -e^{-\fr{e}{i\hbar c} \vp(x)}\fr{i k a_{in}}  {(4\pi)^2} \int_Q    e^{\fr{e}{i\hbar c} \vp(y)}\fr{e^{ik|x-y|}} { |x-y|}(1+\cos\chi)dy.
 \ee
 In particular,
 for  the two-slit diffraction
  we get, similarly to 
(\ref{dif}),
 \be\la{dif2}
 a_\infty^B(x)\sim 
 e^{-\fr{e}{i\hbar c}  [\vp(x)-\vp(q_1)]}
  \fr{e^{ik|x-q_1|}}{|x-q_1|}(1+\cos\chi_1)+
   e^{-\fr{e}{i\hbar c}  [\vp(x)-\vp(q_2)]}
  \fr{e^{ik|x-q_2|}}{|x-q_2|}(1+\cos\chi_2).
 \ee
 \noindent{\bf  The A-B shift.}  Now to determine the maxima of the diffraction amplitude
  $|a_\infty^B(x)|$  we should
  add the phase contribution $ \delta = \fr {e} {\hbar c} [\vp (q_1) - \vp (q_2)] $
  into the equation (\ref {Bragg}),
\be\la{Bragg2}
\sqrt{(x_2-d)^2+D^2}-\sqrt{(x_2+d)^2+D^2}+\de=n\lam,\quad n=0,\pm1,\dots
 \ee
 Accordingly, the Taylor decomposition now gives the equation
  $2d x\approx  [n\lam-\de] D$. 
 This means that  the coordinates of all the maxima are shifted by the same value
  $\de D/(2d)$;
  i.e., the whole diffraction pattern shifts without changing its shape. 
 \br
 {\rm
 i) The phase difference $\vp(q_1)-\vp(q_2)$ is approximately
 $\Phi$ if the distance $|q_1-q_2|$ is sufficiently small.
 \smallskip\\
ii)  
 The limiting amplitude principle (\ref{reap12}) and the Sommerfeld radiation condition 
 (\ref{SRC}) for the limiting amplitudes $a_\infty(x)$ and $a_\infty^B(x)$
 hold   in actual physical situation 
 due to the known results from
 \ci{Eidus69, Eidus89},
 \ci[Ch. 28]{KK2012},
 since the real scatterer is not two-dimensional.
   \smallskip\\
iii)   The formulas  (\ref{reap52}),  (\ref{dif2}) give the solution in the zero order 
approximation in the magnetic field. 
Respectively, the A-B shift holds in this approximation.
To calculate higher order approximations, one should modify the Cauchy data (\ref{Cd0})
using the series (\ref{S11}). Then the diffraction pattern 
in presence of the magnetic field $B$
provisionally does not reduce to a shift of the pattern corresponding to $B=0$.

}
 \er



 \appendix

\setcounter{section}{0}
\setcounter{equation}{0}
\protect\renewcommand{\theequation}{\thesection.\arabic{equation}}
\protect\renewcommand{\thesubsection}{\thesection.\arabic{subsection}}
\protect\renewcommand{\thetheorem}{\Alph{section}.\arabic{theorem}}

\section{Old Quantum Mechanics} \la{s12}
 

We recall  the main achievements
of the classical atomic Lorentz--Thompson theory of 1897--1913 and of 
``Old Quantum Mechanics" (1913--1925), as
developed by Planck, Einstein, Bohr, Debye, Sommerfeld, Pauli, and many others \ci {Born1951, Somm1}. In particular, we recall the introduction of the Bohr--Sommerfeld quantization rules,
selection rules and the theory of normal Zeeman effect. 
This is necessary for the introduction of the Schr\"odinger Quantum Mechanics
and of the Pauli theory for the electron spin.

 \subsection {The Thompson electron and the Lorentz classical theory of the atom} \la {s11}
The existence of electric currents in atoms was predicted by Amper in 1820.
In 1855, Geissler and Pl\"ucker discovered  ``cathode rays" in a vacuum tube. In 1895, Perrin experimentally proved
that these rays carry a negative charge.
In 1893--1897, Thompson conducted a series of 
ingenious
experiments on the deflection of these rays in electrical and magnetic
fields, and came to the conclusion that the cathode rays consist of negatively charged particles -- electrons -- 
that obey the Lorentz equation
\be \la {L}
\cm \ddot x (t) = e [\bE ^ {\rm ext} (x (t), t) + \fr1c \dot x (t) \wedge \bB^\ext (x (t), t) ],
\ee
where $ e <0 $ is the electron charge, $ \cm $ is its mass, and $ \bE ^ {\rm ext} (x, t) $ and $ \bB ^ {\rm ext} (x, t) $ are the  Maxwell external fields.
In these experiments, Thompson for the first time measured  with great accuracy the ratio $ e / \cm $, which enters  equation (\ref {L}). For these experiments, Thompson was awarded the Nobel Prize in 1906.

Equation (\ref {L}) was applied to   atomic electron in a constant uniform magnetic field $ B $, 
replacing
the first term on the right  by an elastic force: 
\be \la {Lat}
\cm [\ddot x (t) + \om_0 ^ 2 x (t)] = \fr ec \dot x (t) \wedge B.
\ee
This {\it linear equation} led to a remarkable explanation of basic atomic phenomena (polarization, dispersion and others), and
served as a basis for the Drude
classical theory of metals  (1900).

In 1897, Zeeman discovered a splitting of  spectral
lines of atoms in magnetic fields.
Part of these experimental
results (``normal Zeeman effect") was 
explained in 1897 by Lorentz based on equation (\ref {Lat}).
This equation  
results in the {\it Larmor precession} of the electron orbit 
around the direction of the magnetic field.
This linear equation (\ref {Lat}) is easy to solve. Namely, choosing $ B = (0,0, B_3) $, we get
\be \label {CA-y-eq}
\ddot x_1 + \omega_0 ^ 2 x_1 = 2 \om_L \dot x_2,
\qquad
\ddot x_2 + \omega_0 ^ 2 x_2 = - 2 \om_L \dot x_1,
\qquad
\ddot x_3 + \omega_0 ^ 2 x_3 = 0,
\ee
where 
$ \om_L$
is the  {\it  Larmor frequency} (\ref{Larf}).
Hence,
$ x_3 (t) = \sin (\omega_0 t - \phi_0) $, and for
 $ z (t): = x_1 (t) + i x_2 (t) $, we get the equation
\begin {equation} \la {comf}
\ddot z + \omega_0 ^ 2 z = -2i \om_L \dot z
\end {equation}
The substitution $ z = Сe ^ {i \omega t} $ gives the characteristic equation
$ - \omega ^ 2 + \omega_0 ^ 2 = 2 \om_L \omega $, whence
\begin {equation} \la {ZL}
 \quad \omega_\pm = - \om_L \pm \sqrt {\omega_0 ^ 2 + \om_L ^ 2} \sim - \om_L \pm \om_0,
 \qquad
 B_3 \to 0.
\end {equation}
Therefore, for small $ B_3 $ we get two solutions $ z (t) = e ^ {i \omega_ \pm t} $
with frequencies $ | \omega_ \pm | \sim \om_0 \pm \om_L $, modified by a magnetic field (we assume that $ \om_0> 0 $).

So, the spectrum of the solution contains 
three different frequencies that make up the {\it triplet}
\be \la {Seeman}
\om_0, \qquad \om_ \pm = \om_0 \pm \om_L,
\ee
which corresponds to the splitting of
the ``atomic  spectral line" $ \om_0 $  into three components. Such splitting was observed 
experimentally 
by Zeeman  in 1895.

This {\it classical} interpretation of atomic radiation turned out to be only asymptotically exact for highly excited atoms,
as it was shown  later by Bohr {\it correspondence principle} (1913), see Section \ref{122}.

Moreover, Lorenz calculated the corresponding
angular distribution of
atomic radiation and its polarization \ci [Section 14.5] {K2013}.
All these calculations  were in a perfect agreement with experimental
observations of Zeeman, and Lorenz together with Zeeman  received the Nobel Prize in 1902.

\subsection {Bohr postulates and Debye's quantization rule} \la {s121}
In 1913, Bohr formulated  postulates (\ref {B1}) and (\ref {B21})
on transitions between quantum stationary orbits.
The question immediately arose of determining the energies of the 
stationary orbits $ E_n=\hbar \om_n $. For the hydrogen atom, the Balmer empirical formula  (1885)
gives with great accuracy the values (\ref {Bal}).
The greatest surprise was caused by the discreteness of the set of possible energies.
In 1911--1913, Ehrenfest conjectured a relation of the discreteness 
with adiabatic invariance of the energy under slow variation of parameters 
\ci [section 52] {A1989}.

In particular, all  basic integrals of dynamical  equations 
are adiabatic invariants: energy, momentum, and angular moment.
Debye in    1913 calculated the action  $ S: = \ds\oint p \, dq $
  for periodic orbits of classical electron in an 
hydrogen atom and discovered that the Balmer formula (\ref {Bal}) is equivalent to the ``quantization rule"
\begin {equation} \la {D}
S = \oint p \, dq
= h n, \quad n = 1,2, \dots, \qquad h = 2 \pi \hbar,
\end {equation}
(see \ci [p. 20] {K2013}), where the integral is taken over the period of the classical electron orbit 
with energy $ E_n $.
Namely, the classical model of the atom is a classical particle (electron), in the Coulomb
nucleous field. This nucleous can be considered fixed,
since its mass is about 1840 times the mass of an electron.
However, all calculations below allow a generalization to the two-body problem, which takes into account the motion of the nucleus. The dynamics of an electron is described
by the 
 Lorentz equation (\ref {L}) with the Coulomb field
$ E ^ {\rm ext} (x) = - eZx / | x | ^ 3 $ (in {\it Gaussian non-rationalized units} cgs)
and $ B = 0 $:
 \be \la {La}
\cm \ddot x (t) = - \fr {e ^ 2Zx (t)} {| x (t) | ^ 3}.
\ee
Multiplying  by $ \dot x $, we obtain the energy integral $ E = \cm \dot x ^ 2/2-e ^ 2Z / | x | $.
Any trajectory lies in a certain plane, for example, in the plane
$ x_3 = 0 $. In particular, for circular trajectories $ x (t) = x_1 + ix_2 = re ^ {- i \om t} $ equation (\ref {La})
gives
 \be \la {La2}
 \cm \om ^ 2r = e ^ 2Z / r ^ 2,
 \ee
and the energy integral turns into $ E = \cm \om ^ 2r ^ 2/2-e ^ 2Z / r $. Eliminating
 $ \om $, we get
 \be \la {La3}
 r = - \fr {e ^ 2Z} {2E}, \qquad
 | \om | r = \sqrt {e ^ 2 Z / (\cm r)} = \sqrt {-2E / \cm},
 \ee
which inplies that
 $ E <0 $.
Hence, for a circular orbit with  period  $ T = 2 \pi / | \om | $
the  action reads
 \begin {equation} \la {DS}
S = \oint p \, dq
= \cm \int_0 ^ T \dot x ^ 2 (t) dt = 2 \pi \cm | \om | r ^ 2 = -2 \pi \cm \sqrt {-2E / \cm} \fr {e ^ 2Z} {2E},
\end {equation}
since
 $ p: = \pa _ {\dot x} \Lam = \cm x \dot x = -i \om \cm re ^ {- i \om t} $, where $ \Lam = \cm \dot x ^ 2 / 2 + e ^ 2Zx / | x | $ is  the Lagrangian of the Lorentz equation (\ref {La}). Therefore, (\ref {D}) is equivalent to
 \be \la {RID}
 e ^ 2Z
 \sqrt {- \fr \cm {2 E_n}} = \hbar n,
 \ee
which coincides with the Balmer formula (\ref {Bal}) with the Rydberg constant (\ref {RID2}),
as in the later Schr\"odinger theory.
This constant, as calculated from available values, was in an excellent agreement
with experimental data, which was a  genuine confirmation
of the Bohr--Debye theory.

\subsection {Correspondence Principle and Selection Rules} \la {122}

In a general form, the  Correspondence Principle was formulated by Bohr in 1920, although he applied it  in various aspects
since 1913. For example: {\it the  frequency of  atomic emission (\ref {B21}) for large quantum numbers $ n, n '$ should go
into the frequency of revolution of classical electron}. 

This  frequency of revolution
was identified until 1913 with the frequency of atomic radiation as in the Lorentz theory
of normal Zeeman effect in Section \ref{s11}. 
The revolution frequency in circular orbits
is calculated from the Lorentz equation (\ref {La}).
Namely, eliminating $ R $ and $ E = E_n $ from (\ref {La3}) and (\ref {RID}), we get
the {\it classic frequency}
\be \la {omcl}
\om_ {c} = \fr {\cm e ^ 4 Z ^ 2} {\hbar ^ 3 n ^ 3}.
\ee
On the other hand, quantum radiation frequencies (\ref {B21}),
with Balmer's terms (\ref {Bal}),
for large $ n, n '$ and bounded differences $ n'-n $ admit
the asymptotics
\be \la {omq}
\om_q = - \fr {2 \pi c R} {n '^ 2} + \fr {2 \pi c R} {n ^ 2}
\sim
\fr {me ^ 4Z ^ 2} {\hbar ^ 3} \fr {n'-n} {n ^ 3}
\ee
by (\ref {RID2}). The minimal frequencies correspond to $ n-n '= \pm 1 $, and 
all other frequencies (called ``obertones") are multiples to the minimal one.
On the other hand, the classical frequency (\ref {omcl}) has no obertones. Therefore, the
correspondence  principle dictates the selection rule
\be \la {prot}
n-n '= \pm 1
\ee
for {\it large quantum numbers} $ n $.

\subsection {Bohr--Sommerfeld Quantization} \la {s123}
 In 1915--1916, Sommerfeld and Wilson suggested to extend the Debye quantization rule
(\ref{D})
  to solutions of
  Hamiltonian systems with several ``periodic"
  degrees of freedom $ q_k $ and corresponding
  canonically conjugate momenta $ p_k $.
   In this case, the suggested
 quantization rules read
 \begin {equation} \la {D1}
\oint p_k \, dq_k
= h n_k, \quad n_k = 0, \pm 1, \dots
\end {equation}

In particular, for  three-dimensional classical atomic system with the Coulomb potential $ -eZ / | x | $
   the Lorentz equation (\ref {L}) takes the form
   \be \la {LL}
\cm \ddot x (t) = - \fr {e ^ 2Zx (t)} {| x (t) | ^ 3} = - \fr {e ^ 2Zx (t)} {| x (t) | ^ 3}.
\ee
The corresponding Lagrange and Hamilton functions are
 \be \la {H3}
\Lam (x, \dot x) = \fr {\cm \dot x ^ 2} 2+ \fr {e ^ 2Z} {| x |}, \qquad
 \cH (x, p) = p \dot x- \Lam = \fr {p ^ 2} {2 \cm} - \fr {e ^ 2Z} {| x |}, \quad p:= \pa_ { \dot x} \Lam = \cm \dot x.
    \ee
Any trajectory of $ x (t) $ lies in some plane $ S $
passing through the nucleus. Therefore, 
we can assume that this equation reads $ \theta = 0 $
in appropriate 
spherical coordinates $ r $ (radius), $ \vp $ (longitude), $ \theta $ (latitude)
with the origine in the nucleus. 
Each bounded trajectory is periodic, and the
quantization rules (\ref {D1}) take the form of integrals over a period
\begin {equation} \la {D2}
\oint p_r \, dr = h n_r, \quad \oint p_ \vp \, d \vp = h n_ \vp, \quad \oint p_ \theta \, d \theta = h n_ \theta = 0,
\end {equation}
because $ d \theta = 0 $.

Let us show that
these quantization rules imply 
the Debye condition (\ref {D}) and, respectively, the Balmer formula (\ref {Bal}) holds for the energies.
First, in the polar coordinates $ r, \vp $ on the plane $ \theta = 0 $, the Lagrangian
reads as
$ \Lam = \cm (\dot r ^ 2 + r ^ 2 \dot \vp ^ 2) / 2 + e ^ 2Z / r $, which implies
$ p_r: = \pa _ {\dot r} \Lam = m \dot r $ and
$ p_ \vp: = \pa _ {\dot \vp} \Lam = m r ^ 2 \dot \vp = \const $ since $ \dot p_ \vp =
   \pa \Lam_ \vp = 0 $. Therefore,
   \be \la {pfi}
 \ds \oint p_ \vp d \vp = 2 \pi p_ \vp = hn_ \vp, \qquad p_ \vp = \hbar n_ \vp, \quad n_ \vp = 0, \pm 1, \dots
   \ee
The Euler--Lagrange equation for $ p_r $ gives
    \be \la {ELr}
    \cm \ddot r = \cm r \dot \vp ^ 2- \fr {e ^ 2Z} {r ^ 2} = \fr {p ^ 2_ \vp} {\cm r ^ 3} - \fr {e ^ 2Z} {r ^ 2},
    \ee
whence integration yields $ \ds \fr {\cm \dot r ^ 2} 2 = - \fr {p_ \vp ^ 2} {2 \cm r ^ 2} + \fr {e ^ 2Z} r + E $.
It is easy to verify that $ E = \cH $; i.e., $ E $ is the energy of the trajectory. Hence,
        \be \la {ELr2}
p_r = \cm \dot r = \pm \fr {\sqrt {2 \cm [E r ^ 2 + e ^ 2Zr] -p_ \vp ^ 2}} r.
    \ee
The minimal and maximal values of the radius $ r_ \pm $ on the trajectory
are obtained from the condition $ \dot r = 0 $. 
The energy $ E  $ should be negative,
since, for bounded trajectories,
the radical expression
should be non-negative only on a finite interval $ [r _-, r _ +] $.
Moreover, $ p_rdr \ge 0 $ according to (\ref {ELr2}). Therefore, 
by integrating \ci [c. 262] {K2013}
the first quantization condition (\ref {D2}) implies
     \be \la {ELr3}
0 \le \oint p_r dr = 2 \int_ {r _-} ^ {r _ +} \ds \sqrt {2 \cm [E r ^ 2 + e ^ 2Zr] -p_ \vp ^ 2}
\fr {dr} r
= 2 \pi [\sqrt {\fr {\cm e ^ 4} {2 | E |}} - | p_ \vp |] = h n_r.
    \ee
Hence, $ \ds \sqrt {\fr {\cm e ^ 4} {2 | E |}} = \hbar (n_r + | n_ \vp |) = \hbar n $ due to (\ref {pfi}).
Therefore,
        \be \la {ELr4}
E = - | E | = - \fr {\cm e ^ 4} {2 \hbar ^ 2n ^ 2}, \qquad n = 1, \dots,
    \ee
which coincides with (\ref {Bal}), (\ref {RID2}). Obviously
$ | p_ \vp | = m r ^ 2 | \dot \vp | $ coincides with
the magnitude of the ``angular momentum vector" $ L: = x \wedge p $, and so
     \be \la {ELr5}
  | L | = p_ \vp = \hbar | n_ \vp |, \qquad 0 \le | n_ \vp | \le n.
    \ee
Finally, note that the Debye quantization condition (\ref {D}) also holds if $ pdq $
define as the invariant
canonical differential form $ pdq: = \sum_k p_kdq_k = p_rdr + p_vpd \vp + p_ \theta d \theta $, since 
in this case 
 $ \ds \oint pdq = \hbar (n_r + n_ \vp) = \hbar n $.
 
 
 \subsection {Atom in a magnetic field} \la {s124}
The Bohr--Sommerfeld quantization (\ref {D1})
can also be applied to the classical
three-dimensional model of  atom
with the Coulomb potential of the nucleus
 in a uniform magnetic field $ B $. Namely,
   the Lorentz equation (\ref {L}) now takes the form
   \be \la {LLB}
\cm \ddot x (t) = - \fr {e ^ 2Zx (t)} {| x (t) | ^ 3} + \fr ec \dot x (t) \wedge B = - \fr {e ^ 2Zx (t)} {| x (t) | ^ 3} + \cm \ti B \dot x (t),
\ee
where $ \ti B v: = \fr e {\cm c} v \wedge B $. The corresponding Lagrange and Hamilton functions read as \ci [(12.81), (12.90)] {K2013}:
 \be \la {H3B}
\Lam_B (x, \dot x) = \fr {\cm \dot x ^ 2} 2+ \fr {e ^ 2Z} {| x |} + \fr ec \dot x \cdot A (x), \qquad
 \cH_B (x, p) = p \dot x- \Lam = \fr 1 {2 \cm} [p- \fr ec A (x)] ^ 2- \fr {e ^ 2Z} {| x |} ,
    \ee
where the vector potential is $ A (x) = \fr12 B \wedge x $, so that $ B = \rot A (x) $, and the momentum is \ci [(12.82)] {K2013}
  \be \la {imp}
  p: = \pa _ {\dot x} \Lam_B = \cm \dot x + \fr ec A (x).
  \ee
The homogeneous
magnetic field can be eliminated by transition to the ``rotating coordinates" $ x'(t) = e ^ {\ti Bt} x (t) $: the trajectory of $ x' (t) $ obviously satisfies the equation (\ref {LLB}) with the
 magnetic field $ B '= 0 $
(``Larmor's theorem").
Hence, the Lagrangian in the rotating frame reads
 \be \la {H3'}
 \Lam'(x', \dot x') = \fr {\cm |\dot x' |^2} 2+ \fr {e ^ 2Z} {| x '|},
  \ee
which coincides with (\ref {H3}). Respectively,
any trajectory $ x '(t) $ lies in some plane of $ S' $
passing through the nucleus, and has the equation $ \theta '= 0 $ in appropriate 
spherical coordinates $ r '$ (radius), $ \vp' $ (longitude), $ \theta '$ (latitude)
with origin at the nucleus. Due to the spherical symmetry of the Coulomb potential, the corresponding vector of angular momentum is conserved.
\be \la {L1}
 L ': = x' (t) \wedge p '(t), \qquad p' (t) = \pa _ {\dot x '} \Lam' = \cm \dot x '(t)
\ee
according to (\ref {H3'}).
In particular, $ L '\bot S' $.
Each bounded trajectory is periodic, and the
 quantization conditions (\ref {D1}) take the form of integrals over the period
\begin {equation} \la {D2'}
\oint p '_ {r'} \, dr '= h n' _ {r '}, \quad \oint p' _ {\vp '} \, d \vp' = h n '_ {\vp' }, \quad \oint p '_ {\theta'} \, d \theta '= h n' _ {\theta '} = 0,
\end {equation}
because $ d \theta '= 0 $.
Here  $ n '_ {r'}, n '_ {\vp'}, n '_ {\theta'} = 0, \pm 1, \dots $.
As was shown above,
these quantization conditions imply 
formulas (\ref {ELr4}) and (\ref {ELr5}):
 \be \la {ELr6}
E'_n = - \fr {\cm e ^ 4} {2 \hbar ^ 2n ^ 2}, \quad n = 1, \dots; \qquad
     | L '| = | p' _ {\vp '} | = \hbar | n' _ {\vp '} |, \quad | n' _ {\vp '} | \le n.
    \ee
Respectively, the
trajectory $ x (t)= e ^ {- \ti Bt} x '(t) $
lies in the rotating plane $ e ^ {- \ti B t} S '$,
and generally  is not periodic, since
the conservation of the angular moment $ L '$ implies
the {\it precession of the angular momentum} around the direction of the magnetic field $ B $,
 \be \la {prec}
 L (t): = x (t) \wedge p (t) = e ^ {- \ti Bt} L '
 \ee
On the other hand, the orbit of $ x (t) $ is periodic {\it in the limit $ B \to 0 $}.
Therefore, Sommerfeld {\it generalizes the quantization conditions} (\ref {D1}) to this case
 \begin {equation} \la {D3}
\oint p_r \, dr \sim h n_r, \quad \oint p_ \vp \, d \vp \sim h n_ \vp, \quad \oint p_ \theta \, d \theta \sim h n_ \theta,
\end {equation}
where $ n_r, n_ \vp, n_ \theta = 0, \pm 1, \dots $.
 \smallskip \\
Sommerfeld derived the following important identity \ci [Chapter II, Section 8, formula (4)] {Somm1}
\be \la {Stoz}
  n_r + n_ \vp + n_ \theta = ne_ {r '} + ne _ {\vp'} + ne _ {\theta '},
 \ee
which follows from the
invariance of integrals of the canonical form $ p\,dq $ along closed curves
with respect to the canonical maps \ci [Section 44] {A1989}. Namely,
the transition from
canonical 
variables $ r, p_r, \vp, p_ \vp, \theta, p_ \theta $ to $ r ', p' _ {r '}, \vp', p '_ {\vp'}, \theta ', p '_ {\theta'} $
conserves the canonical form $ dp \wedge dq $, and therefore,
\be\la{canon} 
  \oint [p_r\,dr+ p_\vp\,d\vp+  p_\theta\,d\theta]
 =
 \oint [p'_{r'}\,dr'+  p'_{\vp'}\,d\vp'+   p'_{\theta'}\,d\theta'],
 \ee
which implies (\ref{Stoz}) according to the quantization rules
(\ref{D2'}) and  (\ref{D3}).
Hence,
  \be\la{Stoz3}
   n_\vp  + n_\theta= n'_{\vp'},
 \ee
(see \ci [Chapter II, Section 8, formula (3)] {Somm1})
since $ n '_ {\theta'} = 0 $, and $ n_r = n '_ {r'} $,
because
$ p_r = p '_ {r'} $. Now the conditions (\ref {ELr6}) with $ B \to 0 $ can be rewritten as
   \be \la {ELr62}
\left \{\ba {rcl}
E_n & = & - \ds \fr {me ^ 4} {2 \hbar ^ 2n ^ 2}, \quad n = 1, \dots
\\
\\
     | L | & = & | p_ \vp | = \hbar | n '_ {\vp'} |, \quad | n '_ {\vp'} | \le n = 1, \dots
    \ea \right |.
    \ee
The quantization rules (\ref {D2}), (\ref {D3}) also imply
the fundamental relation \ci[Chapter II, Section 8, formula (5)]{Somm1})
  \be \la {Sre}
 n_ \vp = n '_ {\vp'} \cos \al,
 \ee
 where $ \al $ is the angle between the vectors $ L '$ and $ B $, which is also equal to the angle
 between the vectors $ L $ and $ B $ due to the precession (\ref {prec}).
This relation  follows from the formula
   \be \la {Sre2}
 p_ \vp = p '_ {\vp'} \cos \al,
 \ee
which is obtained by computing in spherical coordinates. Indeed,
  \be \la {Sre3}
\dot x ^ 2 = \dot r ^ 2 + r ^ 2 \cos ^ 2 \! \theta \, \dot \vp ^ 2 + r ^ 2 \dot \theta ^ 2, \qquad p_ \vp: = \pa _ {\dot \vp} \Lam_B = mr ^ 2 \cos ^ 2 \! \theta \, \dot \vp, \qquad B \sim 0.
 \ee
Similarly, $ p '_ {\vp'} \! \!: = \pa _ {\dot \vp '} \Lam' = mr ^ 2 \cos ^ 2 \! \Theta '\, \dot \vp' = mr ^ 2 \dot \vp '$ because
$ \theta '\equiv 0 $. Both momenta $ p_ \vp $ and $ p '_ {\vp'} $ are conserved, and
 $ \dot \vp = \dot \vp '/ \cos \al $ at maximal latitude
$ \theta = \pi / 2- \al $, which implies (\ref {Sre2}).
Finally, (\ref {Sre2}) and (\ref {D2}), (\ref {D3}) imply (\ref {Sre}) due to the conservation of $ p_ \vp $ and $ p '_ {\vp' } $.
\smallskip \\
\bc
(``Spatial Quantization".)
The angle $ \al $ is quantized by (\ref {Sre}).
Precession (\ref {prec}) implies that $ | L | = | L '| $, and the projection of
vector $ L $ onto  direction of the magnetic field equals $ L_B = | L | \cos \al $. Moreover,
$ | L '| = mr ^ 2 | \dot \vp' | = | p '_ {\vp'} | = \hbar | n '_ {\vp'} | $ according to (\ref {ELr6}).
Therefore,  (\ref {Sre}) implies that $ L_B = \pm \hbar n_ \vp $, where $ | n_ \vp | \le | n '_ {\vp'} | $.
So, denoting $ n '_ {\vp'} = l $ and $ \pm n_ \vp = m $, we get from (\ref {ELr62}) that
  \be \la {Lm}
 | L | = \hbar l, \qquad L_B = \hbar m, \qquad l = 0,1, \dots, n, \quad m = -l, -l + 1, \dots, l.
 \ee
\ec
 \bc \la {cnlm}
Quantum stationary states of an electron in  atom are numbered with three quantum numbers $ n, l, m $
({\it main, azimuth} and {\it magnetic} quantum number, respectively),
and their energy,  angular momentum and its projection onto the direction of the magnetic field are given
by  formulas (\ref {ELr4}) and (\ref {Lm}).
  \ec

\br
{\rm
Quantized values (\ref {Lm}) were obtained for ``infinitely small" magnetic field
$B\to 0$. However, the same expressions should hold also for finite small magnetic field,
since the values of quantum numbers $ n, l, m $ are discrete, while
the action integrals  (\ref {D2}) are {\it adiabatic invariants}. 
}
\er
  
 
\subsection {Normal Zeeman effect} \la {s125}
In 1916, Sommerfeld
obtained from the quantization rules (\ref {Lm}) the same results
for the normal Zeeman effect,
as Lorentz from equation (\ref {Lat}).
Namely,
the classical angular and magnetic moments of the electron are determined by the formulas
\be \la {LM1}
L: = x \wedge p, \qquad M: = \fr e {2 \cm c} x \wedge p.
\ee
Therefore
 \be \la {MLe}
 M = \fr {e} {2 \cm c} L.
\ee
The classical energy (\ref {H3B}) of an electron can be expanded as
\be \la {ecexp}
\cH_B (x, p) = p \dot x- \Lam = \fr {p ^ 2} {2 \cm} - \fr {e ^ 2Z} {| x |}
- \fr e {\cm c} p \cdot A (x) + \fr {e ^ 2} {c ^ 2} A ^ 2 (x) \approx \cH_0 (x, p) - \fr e {\cm c} p \cdot A (x), \quad B \to 0.
\ee
Thus, 
in a magnetic field the classical energy acquires
an additional  amount
\be \la {MH}
- \fr e {\cm c} p \cdot A (x) = - \fr e {\cm c} p \cdot [\fr12 B \wedge x] = - \fr e {2 \cm c} B \cdot [x \wedge p] = - B \cdot M= - \om_L  L_B,
\ee
where $ \om_L$ is the Larmor frequency (\ref{Larf}).
Hence, the total classical energy (\ref{ecexp}) reads
\be \la {MH22}
\cH_B (x, p)
\approx
 \cH_0 (x, p)
- B \cdot M =
\cH_0 (x, p) - \om_L L_B.
\ee 
The corresponding quantum version follows by 
substitution here expression (\ref {Bal}) instead of  the classical energy $\cH_0 (x, p)$ and 
(\ref {Lm}) for $L_B$. 
As a result, we get the following possible values of energy, of
angular momentum and of its projection on the direction of the magnetic field
\be \la {ELr42}
\left \{
\ba {l}
E = E_ {nm} = \hbar\om_n - \om_L \hbar m
\\
\\
L = \hbar l, \qquad L_B = \hbar m
\ea \right | \, \, \, \,
\quad n = 1, \dots; \qquad l = 0,1, \dots, n; \qquad m = -l, \dots, l,
\ee
which  coincides with 
results of the Schr\"odinger theory
(\ref{nlmE2})  except for the range of values for $l$.
Hence, the spectral lines (\ref {B21})
aсquire an additional amount  $- \om_L (m'-m)$. So, the spectral lines in the magnetic field are given by
\be \la {See2}
\om_{nn'}- \om_L  (m'-m)
\ee
which
coincides with  the result of the Schr\"odinger theory (\ref{See}).
Finally,
Bohr's correspondence principle dictates
{\it selection rules},
similar to (\ref {prot}),
\be \la {prot2}
m'-m = 0, \pm 1,
\ee
see
\ci [Supplements 7 and 8] {Somm1}). As a result,
each spectral line (\ref {B21}) becomes the   {\it Zeeman triplet} (\ref {Zeemantr}).
\br
{\rm
The selection rules (\ref {prot2}) also hold in Schr\"odinger's theory, see Section \ref {s7S}.
}
\er

\subsection {Bohr--Pauli theory of periodic table} \la {BPM}

The Bohr and Pauli  theory of periodic table of elements (1921--1923)
was the highest achievement of the ``Old Quantum Mechanics". The theory relies
 on the following postulates:
\smallskip

1) Electrons in many-electrons atoms
are weakly coupled, so that their stationary states
can be described separately by the 
Bohr-Sommerfeld quantization rules.

2) 
Stationary orbits of each 
electron
are numbered by four quantum numbers $ n, l, m, s $, where $ s = \pm 1 $.
Energy, angular momentum and its projection on the direction of the magnetic field are determined by
formulas
(\ref {ELr4}), (\ref {Lm}).
However, now 
{\it the value $ l = n $ is excluded}:
  \be \la {nlms}
  n = 1,2, \dots, \qquad l = 1, \dots, n-1, \qquad m = -l, \dots, l, \qquad s = \pm 1.
  \ee
  
3) The sets of quantum numbers $ n, l, m, s $ are different for different electrons
in accordance with
{\it  the Pauli exclusion principle}.

4) Electrons with the same energy belong to the same shell: the shell $ K $ corresponds to $ n = 1 $, $ L $ corresponds to $ n = 2 $, $ M $ -- to $ n = 3 $, $ N $ -- to $ n = 4 $, ... 
The  atomic shells were introduced by  Mosley 
in his
interpretation of
scattering of
 X-rays by atoms (1913).
\smallskip

The bound $ l \le n-1 $ was found empirically in the ``Old Quantum Theory",
and 
it holds automatically  in the
Schr\"odinger theory by (\ref {nlmE}).
This bound, together with the {\it Pauli exclusion principle} 3), leads to
wonderful
explanation of the periods
in the  table of elements.

For example,
the ground state of the atom corresponds to the minimal energy of the electron configuration.
Hence, in the ground state the electrons 
should belong to shells with possible minimal values of $n$.

The Pauli exclusion principle implies that 
for $ Z \ge 1 $ the shell $ K $ can contain at most  two
electrons with the least energy corresponding to quantum numbers $n=1$, 
$ l = m = 0 $ and $ s = \pm 1 $.

For $ Z \ge 3 $, the shell $L$
can contain at most  $ 8 $ electrons with  
$ n = 2 $, $ l = 0, \dots, n-1 = 1 $, $ m = -l, \dots, l $ and $ s = \pm 1 $.

Similarly, for a shell with any number $ n \ge 1 $, the maximal number of electrons is $ N_n =
2 \sum_ {l = 0} ^ {n-1} (2l + 1) = 2n ^ 2 $.
 These {\it accupation numbers} 
coincide with the famous ``Kabbalistic
sequence" $ 2,8,18,32,50, ... $ of lengths of periods in the table of chemical elements.
For this  exact reason Pauli  introduced in 1923 
the fourth two-digit quantum number $ s = \pm 1 $. 
Otherwise, all periods of the table would turn out to be two times shorter than necessary.

\subsection {Hamilton--Jacobi equation and Optico-Mechanical Analogy} \la {s126}
After 1915 the quantisation rules (\ref {D}) and (\ref {D1})
attracted 
common attention
to the study of  action
and to its calculation using the Hamilton--Jacobi equation (\ref {HJ}).
Many attempts were made to derive a new electron dynamics
from this equation using Hamilton's optico-mechanical analogy (1840),
which is based on the parallelizm berween the Fermat and Maupertuis  variation principles
\ci [p. 246] {A1989}.
This idea was first realized by Schr\"odinger in 1926, who identified the role of the Hamilton--Jacobi equation
with the role of the eikonal equation in optics as the equation for the phase function of shortwave solutions.
The Schr\"odinger equation (\ref {S})  {\bf formally} follows
from 
optico-mechanical 
 analogy \ci {Fermi1961}, and exactly this analogy
was in the center of 
Schr\"odinger's first papers on quantum mechanics \ci [II] {Schr1926} and of his
Nobel lecture \ci {Schr1933}.




\newpage
\noindent {\it Faculty of Mathematics, University of Vienna,  Oskar-Morgenstern-Platz 1, 1090 Wien, Austria, \\Институт проблем передачи информации РАН, Москва 127994, Россия}
\vskip 0.3cm
\noindent {\it e-mail}: alexander.komech@univie.ac.at
\vskip 0.5cm

\end{document}